\documentclass[12pt,preprint]{aastex}
\usepackage{epsfig}
\usepackage{natbib}
\def\plotone#1{\centering \leavevmode
\includegraphics[clip=, width=.75\columnwidth]{#1}}

\newcommand{\cN}[1]{\mathcal{N}}

\def\gsim{\;\rlap{\lower 2.5pt
 \hbox{$\sim$}}\raise 1.5pt\hbox{$>$}\;}
\def\lsim{\;\rlap{\lower 2.5pt
   \hbox{$\sim$}}\raise 1.5pt\hbox{$<$}\;}

\begin{document}


\title{
On Constraining A Transiting Exoplanet's Rotation Rate\\
With Its Transit Spectrum}

\author{David S. Spiegel\altaffilmark{1}, Zolt\'an Haiman\altaffilmark{1},
B. Scott Gaudi\altaffilmark{2}}

\affil{$^1$Department of Astronomy, Columbia University, 550 West 120th Street, New York, NY 10027}

\affil{$^2$Department of Astronomy, Ohio State University, 140 W. 18th Avenue, Columbus Ohio 43210--1173}

\vspace{0.5\baselineskip}

\email{dave@astro.columbia.edu, zoltan@astro.columbia.edu,
  gaudi@astronomy.ohio-state.edu}

\begin{abstract}
We investigate the effect of planetary rotation on the transit
spectrum of an extrasolar giant planet.
During ingress and egress, absorption features arising from the planet's
atmosphere are Doppler shifted by of order the planet's rotational
velocity ($\sim 1-2 {\rm ~km~s^{-1}}$) relative to where they would be
if the planet were not rotating.  We focus in particular on the case of
HD209458b, which ought to be at least as good a target as any other known
transiting planet.  For HD209458b, this shift should give rise to a small
net centroid shift of $\sim 60 {\rm ~cm~s^{-1}}$ on the stellar absorption
lines.  Using a detailed model of the transmission spectrum due to a
rotating star transited by a rotating planet with an isothermal atmosphere,
we simulate the effect of the planet's rotation on the shape
of the spectral lines, and in particular on the magnitude of their width
and centroid shift.  We then use this simulation to determine the
expected signal--to--noise ratio for distinguishing a rotating from a
non--rotating planet, and assess how this S/N scales with
various parameters of HD209458b.  We find that with a 6~m telescope,
an equatorial rotational velocity of $\sim 2 {\rm ~km~s^{-1}}$ could be
detected with a S/N $\sim 5$ by accumulating the signal over many transits
over the course of several years.  With a 30~m telescope, the time required
to make such a detection reduces to less than 2 months.
\end{abstract}


\section{Introduction}
\label{sec:intro}
Since the early 1990s, more than 200 planets have been discovered orbiting
other stars.
Roughly four fifths of the planets that have been discovered are at least half
as massive as Jupiter, and about a quarter of the known planets 
orbit extremely close to their parent star ($\lsim 0.1 {\rm~AU}$).
The Jupiter--mass planets that are in close orbits are highly irradiated by
their stars and are therefore called ``Hot Jupiters.''
Twenty Hot Jupiters transit along the line of sight between Earth and
their star
\citep{charbonneau_et_al2000, henry_et_al2000, konacki_et_al2003,
bouchy_et_al2004, pont_et_al2004, konacki_et_al2004, konacki_et_al2005,
udalski_et_al2002a, udalski_et_al2002b, udalski_et_al2002c, udalski_et_al2003,
udalski_et_al2004, alonso_et_al2004, bouchy_et_al2005, mccullough_et_al2006,
odonovan_et_al2006b, bakos_et_al2006, collier_et_al2006, burke_et_al2007,
gillon_et_al2007, bakos_et_al2007a, bakos_et_al2007bb,
johnskrull_et_al2007}\footnote{Corot--Exo--1b is reported at
{\it http://www.esa.int/esaCP/SEMCKNU681F\_index\_0.html}.}.
Observations of the transiting planets have confirmed
their similarity to Jupiter by revealing that these are gas giant planets,
with radii comparable to, or somewhat larger than Jupiter's
\citep{charbonneau_et_al2000, henry_et_al2000, gaudi2005}.

The effects of the tidal torques experienced by an orbiting body have
been studied for a long time -- for an early seminal analyis, see
\citet{goldreich+peale1966}.  Such torques
tend to synchronize a satellite's rotation rate to its orbital rate, and
if the torque is sufficient this synchronization is achieved and the
orbiter is said to be ``tidally locked,'' as the Earth's Moon is.  The
Hot Jupiter--class extrasolar planets are thought to orbit sufficiently
close to their stars that their tidal locking timescales are much shorter
than the ages of the planets.  The planets, then, are expected to be
tidally locked to the stars, with one hemisphere in permanent day and
the other in permanent night \citep{harrington_et_al2006}.

A tidally locked Hot Jupiter will have a permanent sharp contrast in
temperature between the substellar point and the night side, which must
have a profound influence on the atmospheric dynamics.  
\citet{showman+guillot2002} make simple predictions of the day/night
temperature difference ($\sim 500 {\rm ~K}$) and the speed of winds (up
to $\sim 2 {\rm ~km~s^{-1}}$), and their detailed, three--dimensional
simulations agree with their estimates.  Shallow--water simulations by
\citet{cho_et_al2003} predict longitudinally averaged zonal wind speeds
of up to $400 {\rm~m~s^{-1}}$, with local winds approaching
$2.7 {\rm~km~s^{-1}}$ (under some assumtions).  Simulations by
\citet{cooper+showman2005}
predict a super--rotational jet (i.e., blowing eastward, where north is
defined by the right--hand rule) that blows the hottest part of the planet
downstream by about $60\degr$ from the substellar point.  Their simulations
predict supersonic winds exceeding $9 {\rm~km~s^{-1}}$ at high latitudes,
high in the atmosphere (where the optical depth is low) and winds exceeding
$4 {\rm~km~s^{-1}}$ at pressures near the photosphere.  A Spitzer Space
Telescope phase curve for $\upsilon$ Andromedae b rules out a phase--shift
as large as $60\degr$ between the substellar point and the hottest
spot \citep{harrington_et_al2006}, but a Spitzer phase curve for
HD189733b favors a $\sim 30\degr$ shift for that planet
\citep{knutson_et_al2007b}, so it remains unclear to what extent available
data indicate very strong photospheric winds.

Transmission spectroscopy is a way to probe the atmospheres of these planets.
\citet{charbonneau_et_al2002} were the first to detect an absorption feature
in what is probably the atmosphere of HD209458b, when they found that the
effective radius of the planet increases slightly at the wavelength of a
strong sodium absorption doublet (the sodium D lines) at $\sim 590 {\rm~nm}$.
In addition, \citet{vidal-madjar_et_al2003, vidal-madjar_et_al2004} have
reported a number of absorption features in HD209458's transit spectra that
are due to various species (including hydrogen Lyman alpha, neutral carbon,
oxygen, and sulfur, and some ionization states of carbon, nitrogen, and
silicon) in a hot exosphere that is probably only loosely bound to the planet.
Intriguingly, through analyzing the red and near--IR portion of HD209458b's
transit spectrum \citet{barman2007} found a $10 \sigma$ detection of
atmospheric water vapor.  Several measurements of the planet's emission
spectrum, however, have found results that seem to be inconsistent with
high water abundance high in the atmosphere \citep{grillmair_et_al2007,
richardson_et_al2007, swain_et_al2007}.

Initial work by \citet{seager+sasselov2000} and a comprehensive study by
\citet[hereafter B01]{brown2001}
have described various other considerations that should affect the details of
transit spectra, including the orbital motion of a planet (a few tens of
kilometers per second in the radial
direction), the rotation of the planet (a few kilometers per second at the
equator, according to the hypothesis that the planet is tidally locked), and
winds on the planet's surface (in B01's analysis, up to
$\sim 1 {\rm~km~s^{-1}}$).  These physical effects should tend to broaden or
impose Doppler shifts on absorption features due to the planet's atmosphere.
B01 constructed an impressively detailed model of radiative transfer
through a Hot Jupiter's atmosphere, assuming various models of zonal windflow
superimposed on an equatorial bulk rotation speed of
$v_{\rm eq} = 2 {\rm~km~s^{-1}}$, which is approximately the value for
HD209458b under the assumption that it is tidally locked in its 3.5 day orbit.
He finds the height of the cloud deck to be the most important
parameter that affects the transmission of light through the planet's
atmosphere.

The original discovery of the roughly Jupiter--mass planet in a
close, $\sim 4$ day orbit around 51 Pegasi \citep{mayor+queloz1995} prompted
interest in the dynamics and structure that must govern a highly insolated
gas giant planet \citep{guillot_et_al1996}.  Observations of the transiting
Hot Jupiters heightened this interest when they revealed a puzzling feature of
these planets: at least several of them are a bit puffier than Jupiter, with
diameters ranging from slightly larger than Jupiter's to as much as $\sim 80\%$
larger.  It is not clear what allows some planets to maintain such large radii.
It has been suggested that if a
Jovian planet migrates very quickly, from its presumed formation location at
least several AU from its star, to its eventual several day orbit, then it
might reach its final home before it has cooled enough to shrink to Jupiter's
radius.  Accordingly, some authors have investigated the migration processes
that lead gas giant planets to such close orbits as have been found 
\citep[e.g.][]{trilling_et_al2002}.  Others have investigated various ways
in which a gas giant could either be heated once it ends up near its star, or
otherwise maintain sufficient internal energy to sustain its inflated size
\citep{guillot+showman2002, burrows_et_al2003, laughlin_et_al2005,
bodenheimer_et_al2003, guillot2005, burrows_et_al2007, chabrier+baraffe2007}.
Although various physical mechanisms have been suggested as the apparently
missing energy source that allows the unexpectedly large radii sometimes seen,
the lesson of these investigations {\it in toto} is that it is not easy to
explain the inflated sizes, either in terms of the greater stellar flux that
these planets experience by virtue of being so close to their stars, or in
terms of their evolutionary migratory histories.  A recent paper by
\citet{winn+holman2005} propose that, contrary to
the commonly accepted paradigm, Hot Jupiters might be trapped in a
Cassini state with large obliquity, in which the spin--axis precesses in
resonance with the orbit, but lies nearly in the orbital plane.  Such a state
might be stable against perturbation, and yet able to generate sufficient
internal energy to increase a gas giant planet's radius to the observed
values.  In light of an even more recent analysis by
\citet{levrard_et_al2007}, however, it appears that the probability of
capture into a Cassini state 2 resonance is quite small for a planet with
semi--major axis $a < 0.1 {\rm ~AU}$.  Furthermore, \citet{fabrycky_et_al2007}
argue that even if a planet is captured into Cassini state 2, it is likely to
remain there for a time that is short relative to the age of the system.

High--resolution transit spectra that have high signal--to--noise ratios will
allow us to distinguish between various models of orbit, rotation, and
weather, as discussed by B01.  Because the orbit is known to high
accuracy, and the predictions of the effects of weather (or climate) are
highly uncertain, as described above, we will focus in this paper on the
much more easily predicted effect of a planet's rotation on a
transit--spectrum.  If we neglect winds, then the large--obliquity Cassini
state described by \citet{winn+holman2005} should have a spectral signature
that is very similar to that of a non--rotating model.  In contrast, the
rotation of a tidally locked planet should impose a Doppler distortion
on spectral lines arising from the planet's atmosphere that
is roughly an overall redshift during ingress, as the planet is just
entering the stellar disk, and a similar distortion that is roughly an
overall blueshift during egress, as the planet is just exiting the disk.
During mid--transit, the spectral distortion is more similar to rotational
broadening.  In the present investigation, we address whether there is any
hope that these spectral distortions from tidally--locked rotation can be
observed.  In our study, we focus only on the sodium doublet detected
by \citet{charbonneau_et_al2002}.  As we will show below, the sensitivity
of a measurement of rotation scales with the square root of the number of lines
under consideration.  Model spectra from, e.g., \citet{sudarsky_et_al2003} and
\citet{barman2007} predict a strong potassium doublet at
$\sim 770 {\rm~nm}$, strong water absorption features in the near--infrared,
and a handful of near--UV lines.  If some of these are confirmed in the
atmosphere of a transiting planet, they will provide a modest
increase in S/N.  Since the sodium lines are expected to be the strongest,
however, it seems unlikely that observing multiple lines will yield a
boost in S/N by more than a factor of a few.

We emphasize that it may not be at all justified to neglect winds.  It is
quite likely that there are super--rotational winds on Hot Jupiters, which
are probably necessary to heat the ``night'' side.  As indicated above, some
models predict, and the observed phase curve for HD189733b suggests, that at
the photosphere these winds might be significantly (100\% or more) greater
than the equatorial rotation rate, and therefore might contribute
importantly to the Doppler distortion induced by the motion of the planet's
atmosphere.  Nevertheless, in order to isolate the contribution of rotation,
we do neglect winds in this study.  The Doppler distortions that we predict
can therefore probably be taken as a lower bound on the distortions that would
be observed for a tidally--locked transiting Hot Jupiter.

We find that the spectral shifts induced by rotation will be difficult to
detect with current technology, but perhaps not insurmountably so, at least
with technology that might be available in the not--to--distant future.
The measurements we will describe are limited by a paucity of photons.
As such, their signal--to--noise ratio will be enhanced by a bright star and
a puffy planet (i.e., a planet with a large scale--height).  HD209458 is
at least a magnitude brighter than any other star with a known transiting
planet except HD189733, and its planet is larger than HD189733b; so HD209458b
should be a better target than any other known transiting planet except
possibly HD189733b.  In this paper, we model the HD209458b system because it is
the best--studied system, and it is unlikely that any currently--known
planets would be significantly better targets.
In a single transit, observations of HD209458 with a 6~m
telescope that has a high--resolution ($>50,000$) optical spectrograph
with good throughput ($\sim 18\%$) could only show the influence of tidally
locked rotation at the $\sim 0.2\sigma$ level.  With ultrahigh--resolution
($\gsim 700,000$) and good throughput ($\sim 4\%$) this effect would still
only show up at the $\sim 0.6 \sigma$ level.  In less than a year, the signal
of rotation could be present at five times the noise ($S/N = 5$).  Of
course, a telescope with larger collecting area, higher spectral
resolution, or better throughput would cause the signal to be apparent
at that significance level in less time.

Other studies have approached the problem of determining the rotation
rate from a different angle.  \citet{seager+hui2002} and
\citet{barnes+fortney2003} suggest that an oblate spheroid will have a
different transit light curve from a perfect sphere, and so measuring
the oblateness from transit photometry will provide a handle on the
rotation rate.  The oblateness is somewhat degenerate with several
other parameters that are not perfectly known, however, so they
conclude that it would be difficult to actually determine that
rotation rate in this manner.  The method we describe here could
eventually prove to be an important complement to other observations
to constrain the rotation rate.

In the remainder of this paper, we address this idea in detail.
One complication that we discuss below is that the technique of this paper
is not immune from several near--degeneracies among the many
attributes of transiting extrasolar planets that influence light
curves or spectra.  Although it is likely that current or near--future
instruments will be sensitive enough that the spectral distortion
imposed by HD209458b's rotation (if it is tidally locked) is
visible, it might still be very challenging to discern the fingerprint
of rotation from other attributes that affect the spectra at a similar
level.  In this paper, we tackle the forward problem of calculating the
amount of distortion that is caused by rotation.  The inverse
problem -- determining from observations whether a planet is tidally
locked -- is more difficult and should be the topic of a future study.

The structure of the rest of this paper is as follows:
In \S~\ref{sec:overview}, we describe qualitatively what happens to
the starlight received on Earth when a planet transits its star; we
give a rough order of magnitude estimate of the the magnitude and
detectability of the spectral distortions caused by tidally locked
rotation; and we briefly describe some technological progress and
remaining challenges relevant to our task of deducing bulk motions
in a planet's atmosphere from transit spectra.  In \S~\ref{sec:model},
we describe our computational model of a transit spectrum.
In \S~\ref{sec:results}, we describe the results of our model
according to various assumed input parameters.
In \S~\ref{sec:disc}, we discuss the scaling of S/N on various model
parameters and we address the prospects of actually observationally
determining whether a transitting planet is tidally locked.  In
\S~\ref{sec:conc}, we conclude
by describing various ways to boost our predicted S/N to a more
optimistic value.

\section{Overview of the Problem}
\label{sec:overview}

The practical feasibility of the investigation we undertake depends on a
few factors: understanding the various detailed processes that affect
the starlight that reaches Earth when a planet transits its star; the
magnitude of the distortion that tidally locked rotation induces; and the
technology available to measure such distortions.  In this section, we
give an overview of these three factors -- in particular, in
\S~\ref{ssec:preview}, we give a simple estimate of the results that
we will later (in \S~\ref{sec:results}) calculate in detail.

\subsection{Relevant Processes}
\label{ssec:processes}
\begin{figure}[h!]
\plotone
{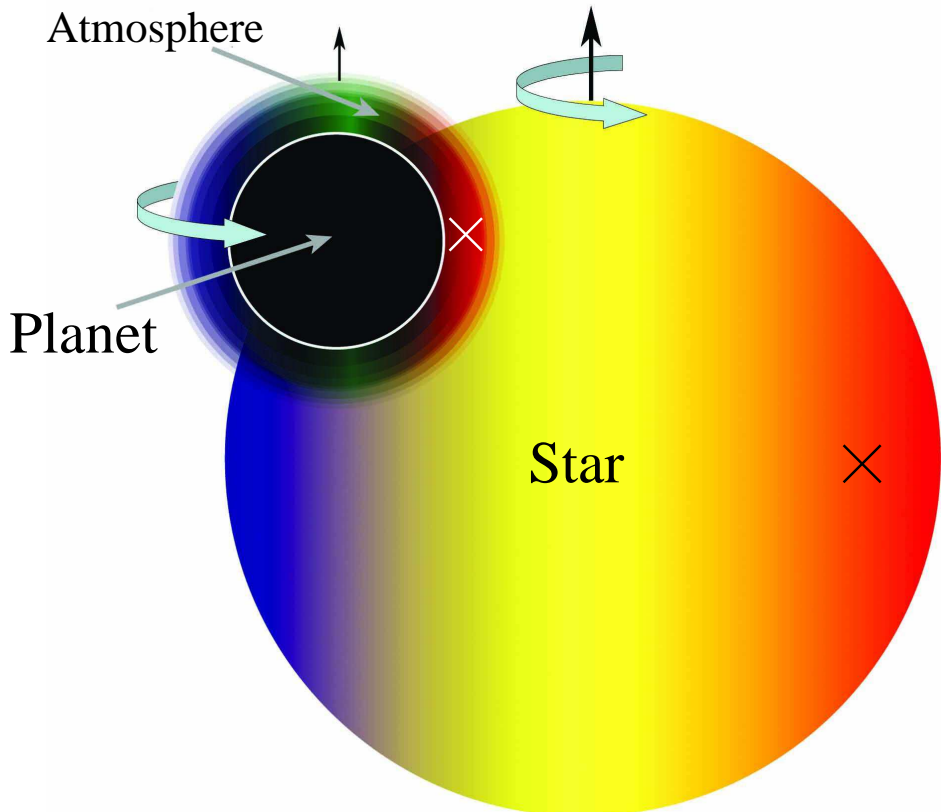}
\caption{Rotating planet beginning to transit in front of rotating star.
The vertical black arrows represent the rotation axes of the planet and the
star, and the curved arrows indicate the direction of rotation for each.
The X's on the right--sides of both the planet and the star indicate regions
that are receding from the observer and are therefore redshifted;
the unmarked left sides of the planet and the star are moving toward the
observer and are therefore blueshifted.
The white circle surrounding the opaque black part of the planet denotes
the cloud deck, or the boundary between the partially transparent and the
fully opaque portions of the planet's disk.  The planet is orbiting in the
same sense as both it and the star are rotating.  The planet is shown above
the star's midplane to represent the inclination of the orbit relative to
the line--of--sight.}
\label{fig:schematic}
\end{figure}

A planet transiting in front of its star affects the starlight that ultimately
reaches Earth in many ways.  The motion of the planet's atmosphere
(rotation and winds) is a small perturbation on top of several more
dominant effects.  We therefore summarize below the physical processes
that are at least as significant as the effect of tidally locked rotation.
Figure~\ref{fig:schematic} schematically represents this situation, and
captures nearly all of the processes described below:
a rotating planet (of exaggerated relative size) transits in front of a
rotating star.  The figure depicts a snapshot partway through ingress, when
half of the planet is in front of the stellar disk\footnote{The planet is
above the star's equator to represent a slight inclination in its orbit.}.
The white circle indicates a hypothetical sharp demarcation between the opaque
part of the planet (in black) and the optically thin part, labeled
``Atmosphere'' (described further below).

\begin{enumerate}
\label{list:transiteffects}
\item {\it Geometric Occultation}:\\
 The largest effect is an overall dimming by a factor of roughly the
ratio of the area of the planet to that of the star: $(R_p/R_*)^2$.  Since
stars are not perfectly uniform disks, but instead tend to darken toward
the limb at most visible wavelengths, the fractional dimming due to being
in the planet's shadow tends to be slightly less than the ratio of the
areas when the planet is near the edge of the stellar disk and slightly
more than this ratio when the planet is near the center.
\item {\it Stellar Wobble}:\\
 The primary spectral effect of the planet orbiting the star is the radial
velocity wobble induced by the planet's gravity.  This periodic spectral
shift is of course in effect during the transit, when, for a close--in planet
like HD209458b, it has an influence on the order of
$\sim \pm 10 {\rm ~m~s^{-1}}$.
This effect is a redshift as the planet begins to transit across the disk
(during ingress) and a blueshift during egress.
\item {\it Rossiter--McLaughlin Effect}:\\
 A more subtle effect arises because, during the transit, the planet moves
across -- and therefore blocks -- parts of the star that have different
recessional velocities.  If (as is expected) the planet's orbit is aligned
with the star's spin, then during ingress the planet is blocking a part of
the star that is slightly blueshifted, and during egress it is blocking a part
of the star that is slightly redshifted.  Figure~\ref{fig:schematic}
illustrates the planet blocking some of the bluest parts of the star
during ingress.

The parts of the star that are occluded during ingress/egress have spectra
that are blue/redshifted by a velocity that is approximately the equatorial
rotational speed of the star, or about $\sim 1$-$2 {\rm ~km~s^{-1}}$ for a
Sun--like star.
As the figure indicates, during ingress/egress, the integrated
spectrum of the remaining (unblocked) parts of the star is on average slightly
redder/bluer than it would be if the planet were entirely transparent.
Therefore, during ingress, the centroids of stellar lines are shifted
slightly to the red, and during egress the centroids are correspondingly
shifted to the blue.

This so--called Rossiter--McLaughlin effect (RME), described originally by
\citet{rossiter1924} and \citet{mclaughlin1924} in the case of eclipsing binary
stars, adds to the shifts already caused by the radial velocity induced by
the planet's gravity, described in (2.) above.  The RME has been described in
depth more recently in the context of extrasolar planets by
\citet{ohta_et_al2005}, \citet{gimenez_et_al2006}, and \citet{gaudi+winn2007}.
These centroid--shifts are expected to be comparable in magnitude to the radial
velocity wobble from the planet's gravity, and can be roughly estimated as
\[
\left| \delta v_{\rm R-M} \right| \sim 1 {\rm ~km~s^{-1}} \times (R_p/R_*)^2 \sim 10 {\rm ~m~s^{-1}},
\]
In fact, the amount of the shift can be predicted precisely for a given
orientation of the planet's orbit, and so measuring the shift is
tantamount to measuring the alignment between the star's spin and the planet's
orbit.  Three years ago, \citet{winn_et_al2005} first found that the spin of
HD209458 and the orbital plane of its planet are nearly aligned.  The degree
of alignment has been measured for two other systems -- \citet{winn_et_al2006}
found that the spin of HD189733 and its planet are also nearly aligned,
and \citet{narita_et_al2007} measured a mis--alignment between these two
vectors by $\sim (30 \pm 20)\degr$ in the TrES--1 system. 
\item {\it Planet's Atmospheric Opacity}:\\
 Furthermore, a gas--giant planet's opacity surely does not have a
perfectly sharp discontinuity at an outer boundary, as a billiard ball does.
Instead, it has an extended atmosphere in which the opacity must vary more or
less smoothly.  There may be a cloud layer, below which the planet is
entirely opaque to tangential rays and above which the opacity varies smoothly.
Most critical to our investigation, at a given radius, the planet's opacity
to tangential lines of sight must vary with wavelength, depending on the
contents of its atmosphere.  At wavelengths of strong atomic or molecular
transitions, the planet's atmosphere will be more opaque than at other
wavelengths.  As a result, the effective radius of the planet, or the
radius at which the optical depth along a tangential ray is of order unity,
is greater at some wavelengths than at others.  These effects have been
described in detail by B01.
\item {\it Planet's Orbital Motion}:\\
 The motion of the planet's atmosphere must influence the transit spectrum in
several delicate ways.  As B01 points out, there are three main mechanisms by
which the motion of a planet's atmosphere relative to its star can affect
the spectrum: the planet's orbital velocity along the line--of--sight, the
planet's (possibly tidally locked) rotation, and winds in its atmosphere.
The largest effect of these three is the orbital velocity, which imposes a
bulk blue/redshift during ingress/egress of $\sim 15 {\rm ~km~s^{-1}}$ to
spectral lines arising from the planet's atmosphere.  These shifts are of
opposite sign to the radial velocity wobble and to the shifts from the RME, and
therefore tend to lessen the apparent RME slightly.
\item {\it Planet's Atmospheric Motion}:\\
 The most dynamically interesting (and subtlest) effects are those caused by
the planetary rotational velocity and atmospheric winds.  Since a tidally
locked planet rotates in the same sense as it orbits, the rotational velocity
of its outside edge has the same sign as its orbital velocity, and the
rotational velocity of its inside edge has the opposite sign.  As a result,
during the beginning of ingress and the end of egress, when only the inside
edge of the planet is over the star, tidally locked rotation will impose a
spectral distortion that is in the opposite sense of that caused by the bulk
orbital velocity described in (5.) above, and that is in the same sense
as the RME: the distortions are roughly equivalent to a relative redshift
during ingress (graphically represented in Figure~\ref{fig:schematic})
and a relative blueshift during egress.
During mid--transit, with some parts of the rotating planet's atmosphere
moving toward and other parts away from the star relative to an otherwise
identical but non--rotating planet, the overall influence of the planet's
rotation is approximately equivalent to rotational broadening.

Winds complicate the picture even further.  It is likely that winds tend
to rush from the substellar hot spot to the colder night side of the planet.
With the substellar point on the opposite side of the planet from Earth
during a transit, this corresponds to winds rushing toward us at several
hundred to several thousand meters per second.  This would tend to
blueshift the spectrum throughout the transit.  Zonal wind bands, somewhat
similar to those on Jupiter but with much higher speeds, or other more detailed
winds, can have an even more intricate effect.
\item {\it Additional Effects}:\\
If a transiting planet were to have nonzero orbital eccentricity, or rings,
these could complicate a measurement of rotation rate.  Nonzero eccentricity
would break the symmetry between ingress and egress.  Still, if the orbit were
well-known, this could be modeled and taken into account.  It seems unlikely
that a Hot Jupiter could maintain rings: Icy rings would sublimate,
and, if not continuously replenished, dusty/rocky rings would quickly
succumb to the Poynting--Robertson effect \citep{poynting1903,robertson1937}.
But if, somehow, a ring were to find a way to persevere around a transiting
Hot Jupiter, it could confound -- perhaps hopelessly -- a measurement of
rotation.  The consequences of rings for the Rossiter--McLaughlin effect is
addressed in \citet{ohta_et_al2006}.  Saturn's rings are nearly four--times the
area of the planet, so for a planet (with equatorial rings that are as
relatively large as Saturn's) whose orbit is tilted an angle 0.1 (in radians)
from edge-on, the rings would be $\sim 40\%$ the area of the planet, which
would increase the RME by $\sim 40\%$.  Uncertainty about the presence and
size of a ring introduces an uncertainty in the size of the RME effect that is
probably larger than the size of the rotation effect.  Furthermore, a ring
would occlude a (small) part of the planet's atmosphere, which would (slightly)
reduce the strength of the rotation signal.
\end{enumerate}

Other interesting phenomena that primarily affect a transit light curve,
rather than the spectrum, include star--spots \citep{silva2003}, atmospheric
lensing \citep{hui+seager2002}, and finite--speed--of--light effects
\citep{loeb2005}.  Although \citet{winn+holman2005} describe a possible
configuration (Cassini state 2) that would produce a spectral signature
that is nearly identical to what would be expected from a non--rotating
planet, the likelihood that any Hot Jupiters are in such a configuration
might be low, and it seems quite likely that {\it some} transiting
planets are not in this state.  Nonetheless, the motion of a transiting
planet's atmosphere -- rotational, wind, or other -- is clearly interesting,
and the basic technique that we describe below is applicable to any model of
atmospheric motion.

\subsection{Preview of Results}
\label{ssec:preview}
A rough estimate of the velocity--shift that is imposed during ingress to the
centroids of the stellar Na D--lines by the planet's tidally locked rotation
(on top of the RME and the shift from the planet's orbital velocity, both of
which would be present even if the planet were not rotating) is the following:
\begin{eqnarray}
\nonumber \delta v & \sim &  \left(\left< \cos[\phi] \right>_{-\pi/2}^{~\pi/2}\right) \times \left( \frac{1}{2} \times \frac{{R_p}^2}{{R_*}^2}\right) \times \left( \frac{2 \pi R_p \Pi_{\rm atm}}{\pi {R_p}^2} \right) \times v_{\rm eq} \\
\nonumber & \sim & 0.64 \times 1\% \times 15\% \times 2000 {\rm ~m~s^{-1}} \\
\label{eq:Delta v estimated} & = & 1.9 {\rm ~m~s^{-1}}.
\end{eqnarray}
In this equation, $\phi$ is a planet--centered azimuthal angle,
$R_p$ and $R_*$ are the planet's and star's radius, respectively,
$\Pi_{\rm atm}$ is the height of the planet's atmosphere, and $v_{\rm eq}$
is the equatorial rotation speed.  The rotation speed at angle $\phi$ is
$v_{\rm eq} \cos[\phi]$.  We take the average of $\cos[\phi]$ from
$-\pi/2$ to $\pi/2$ to get the average planetary rotation speed.
We have used $\Pi_{\rm atm} = 7500 {\rm ~km}$, or 15 times the
presumed scale height of $500 {\rm ~km}$, because the sodium lines
are so heavily saturated that at the assumed abundance and cloud deck
height in our model the line cores do not become optically thin until
that height.  \citet{burrows_et_al2004} and \citet{fortney2005} describe
how the optical depth along tangential rays is greater than the optical depth
on normal rays. The product
\[
\delta_{\rm atm} \approx \left( \frac{1}{2} \times \frac{{R_p}^2}{{R_*}^2}\right) \left( \frac{2 \pi R_p \Pi_{\rm atm}}{\pi {R_p}^2} \right) = \left(\frac{R_p}{R_*}\right)^2 \left( \frac{\Pi_{\rm atm}}{R_p} \right)
\]
is the ratio of the area of the portion of the planet's atmosphere that is in
front of the star halfway through ingress to the total area of the disk of the
star.  Based on this estimate, we expect a maximum velocity shift of
$\delta v \sim 190 {\rm ~cm~s^{-1}}$.  If we take into account that HD209458b's
orbit is actually slightly inclined relative to the line of sight, the cosine
average decreases to $\sim 0.45$, and the total estimate decreases to
$\sim 140 {\rm~cm~s^{-1}}$.  This estimate is in reasonably good agreement with
the centroid--shifts predicted by the full model calculation below
($\sim 60 {\rm ~cm~s^{-1}}$); the
difference between the estimates is most likely due to the difference between
the shapes of the stellar and planetary lines.

We now estimate the signal--to--noise ratio for the detectability of this
effect in an observation of duration $\Delta t$, with a telescope that has
diameter $D$ and throughput efficiency $\eta$.  The signal is the distortion
of the spectrum relative to a non--rotating planet, and for now we will assume
that the noise is
dominated by photon noise.  If a spectrum $F[\lambda]$ with a symmetric
absorption feature of depth $\Delta F$ centered at $\lambda_0$ is redshifted
by an amount $\Delta \lambda$ to
$\widehat{F}[\lambda] \equiv F[\lambda - \Delta \lambda]$, what is the
integrated absolute difference $|F-\widehat{F}|$ over some wavelength range
$2L$ centered on $\lambda_0$?  If the absorption feature is wide compared
with $\Delta \lambda$, then, by symmetry,
\begin{eqnarray}
\nonumber S & = & \int_{\lambda_0 - L}^{\lambda_0 + L} \left| F[\lambda] - F[\lambda - \Delta \lambda] \right| d\lambda \\
\label{eq:signal integral} & \approx & 2 \int_{\lambda_0}^{\lambda_0 + L} \left( F[\lambda] - F[\lambda - \Delta \lambda] \right) d\lambda;
\end{eqnarray}
and if $\Delta \lambda$ is small then
\begin{eqnarray}
\nonumber S & \approx & 2 \Delta \lambda \int_{\lambda_0}^{\lambda_0 + L} F'[\lambda] d\lambda \\
\label{eq:integrated diff} & \approx & 2 (\Delta \lambda) (\Delta F).
\end{eqnarray}

We may now estimate the S/N of our effect (for a single absorption line) using
the lesson of equation~(\ref{eq:integrated diff}), provided we know the
absolute normalization of the stellar spectrum (the number of photons per unit
wavelength).  A spherical blackbody of radius $R_*$ and temperature $T_*$,
at distance $d$ from the telescope, has a photon  number flux at wavelength
$\lambda$ of
\begin{eqnarray}
\nonumber \frac{d \dot{N}_\gamma}{d \lambda} & \sim & B_\lambda[\lambda,T] \left( \frac{1}{h c/\lambda} \right) \left( \frac{\pi {R_*}^2}{d^2} \right) \times \eta \pi (D/2)^2\\
\label{eq:spectrum} & = & \frac{\pi^2 c}{2 \lambda^4 \left( \exp[(h c)/(\lambda k T_*)] -1 \right)} \times \eta \left(\frac{{R_* D}}{d}\right)^2,
\end{eqnarray}
where $B_\lambda$ is the Planck function.
Since the fractional decrease in the spectrum at the line--center is
approximately $\delta_{\rm atm}$, we may express the parameter $\Delta F$
from equation~(\ref{eq:integrated diff}) as
$\Delta F \approx \delta_{\rm atm} (d \dot{N}_\gamma / d \lambda)$.
Similarly, since the root--mean--square velocity shift during ingress is
$\left< v^2 \right>^{1/2} \sim (1/2) \times( 2000 {\rm~m~s^{-1}}) = 1000 {\rm ~m~s^{-1}}$,\footnote{We write $(1/2) \times (2000 {\rm~m~s^{-1}})$ because the
mean value of $\cos^2$ from $-\pi/2$ to $\pi/2$ is $1/2$}
we may express the parameter $\Delta \lambda$ as
$\Delta \lambda \sim (\left< v^2 \right>^{1/2} / c) \times \lambda_0$.
The distortion (the signal) from a single line can therefore be estimated as
\begin{eqnarray}
\nonumber S & = & \delta N_\gamma \sim 2 (\Delta \lambda) \left( \delta_{\rm atm} \frac{d \dot{N}_\gamma}{d \lambda}\right) \Delta t\\
\label{eq:signal estimate} & = & \frac{\pi^2 c \left(\delta_{\rm atm}\right) \Delta \lambda}{\lambda^4 \left( \exp[(h c)/(\lambda k T_*)] -1 \right)} \times \eta \left(\frac{{R_* D}}{d}\right)^2 (\Delta t).
\end{eqnarray}
The shot--noise is the square root of the number of photons in a wavelength
range $2L$ roughly equal to the FWHM of the line, or about $7 {\rm ~km~s^{-1}}$
for a heavily saturated line such as the Na D lines under consideration:
\begin{eqnarray}
\nonumber N & \sim & \sqrt{\frac{d \dot{N}_\gamma}{d\lambda} (2L) (\Delta t)} \\
\label{eq:noise estimate} & \sim & \sqrt{\frac{\pi^2 L c (\Delta t)}{\lambda^4 \left( \exp[(h c)/(\lambda k T_*)] -1 \right)}} \times \sqrt{\eta} \left( \frac{R_* D}{d} \right).
\end{eqnarray}
We estimate the total signal--to--noise ratio arising from a single
absorption line, during an ingress integration of duration $\Delta t$, to be
roughly
\begin{eqnarray}
\nonumber S/N & \sim & \frac{\pi(\delta_{\rm atm})}{\sqrt{\exp[(h c)/(\lambda k T_*)] -1}} \left( \frac{ \Delta \lambda}{\lambda} \right) \left( \sqrt{\frac{c \Delta t}{ L}} \right) \left( \frac{R_* D}{d \lambda} \right) \sqrt{\eta} \\
\nonumber & \sim & \left( 6.6\times 10^{-4} \right) \left( 3.3\times 10^{-6} \right) \left( 2.1\times 10^{11} \right) \left( 5.0\times 10^{-3} \right) \sqrt{\eta} \\
\label{eq:SN estimate} & \sim & 2.3 \sqrt{\eta}.
\end{eqnarray}
The above calculation uses parameters for HD209458 and its planet, a
sodium D line, and a 6~m telescope:
$\lambda = 600 {\rm~nm}$; $\Delta t = 1000 {\rm~s}$;
$R_* = 7.3\times 10^{10} {\rm~cm}$;
$T_* = 6100 {\rm ~K}$, $d = 47 {\rm~pc}$;
and $D = 600 {\rm~cm}$.  For two identical
absorption lines, we gain a factor of $2^{1/2}$ in S/N, and for egress we gain
another factor of $2^{1/2}$, giving a total one--transit S/N of roughly
$4.6 \eta^{1/2}$, not counting the additional signal available during
mid--transit (see further discussion below).  This S/N ratio is in principle
independent of the spectral resolution of the spectrograph, for sufficiently
high spectral resolution.  For low spectral resolution, however, the S/N could
be lower than this estimate (below, we conclude that the S/N loses its
dependence on resolving power for spectral resolution $\gsim$~500,000).

There were several optimistic assumptions that went into this estimate.
Still, this rough estimate of the degree to which a planet's rotation
influences its transit spectrum indicates that the more in--depth study
that we perform below is warranted.

\subsection{Available Technology}
\label{ssec:tech}
Detecting the centroid--shifts caused by tidally locked rotation
($\lsim 1 {\rm ~m~s^{-1}}$) will require very precise measurements of
stellar transit spectra.  Obtaining such high precision spectra will be
quite challenging, for a number of reasons, several of which were
described in the groundbreaking paper by \citet{butler_et_al1996}
that analyzes the limits of Doppler precision.  Of particular concern,
stellar pulsations and turbulent motions in stellar photospheres can cause
small regions of the stellar disk to move at up to $300 {\rm ~m~s^{-1}}$
\citep{dravins1985, ulrich1991}.  These motions tend to average out to produce
stellar spectra that are largely stable; but it is likely that at least some
giant convection cells are not small relative to the size of a planet, and
these could introduce a contaminating source of noise when they are located
behind the planet or its atmosphere.
\citet{butler_et_al1996} reviewed what was then known about the
variability of stellar line--profiles; the upshot is that line--widths may
vary by up to several meters per second over several years, but it is not
clear to what extent spurious apparent velocity shifts may be induced by
convection, and such stellar jitters may prove to be a significant source of
noise that would render it difficult to measure sub meter--per--second
velocity--shifts.  More recently, \citet{bouchy_et_al2005b} have actually
achieved sub meter--per--second accuracy with the HARPS instrument (spectral
resolution of 115,000), and they have found a dispersion in night--averaged
radial velocity measurements for a particular star (HD160691) of
$\sim 0.4 {\rm ~cm~s^{-1}}$ for nights when they took many ($\gsim 200$)
observations.  Since in our situation (taking spectra during ingress, say) we
have minutes, not hours, available, the rms scatter in ingress--averaged radial
velocity measurements is likely to be larger than what they found.  In addition
to the difficulties posed by several systematic sources of noise, achieving
sufficient photon statistics will be difficult for two reasons: for a given
throughput efficiency $\eta$, higher spectral resolution means fewer photons
per bin; and $\eta$ tends to decrease with increasing spectral resolution
$R_S$.

By the mid--1990s, the timeless quest for high--resolution spectrographs
reached a milestone at the Anglo--Australian Telescope with the development
of UHRF and its resolving power of up to $1,000,000$ \citep{diego_et_al1995}.
Despite impressive throughput relative to previous endeavors, however, its
efficiency was insufficient to obtain the sub decameter--per--second Doppler
precision on a $V \ge 7$ star that would be required for planet
searches.  With a $R_S = 600,000$ spectrograph built at Steward Observatory,
\citet{ge_et_al2002} obtained stellar spectra with $R_S \sim 250,000$ and
throughput of 0.8\%.  Furthermore, they predicted that by optimizing their
technology they could increase the throughput to 4\%.  More recently,
\citet{ge_et_al2006} detected a new planet, around HD 102195, with the
Exoplanet Tracker instrument at Kitt Peak.  This instrument has resolution
of $R_S \sim 60,000$ and total throughput of 18\%.
Plans for a spectrograph that has resolving power of 120,000 on a
thirty meter telescope \citep{tokunaga_et_al2006} give cause for optimism
that increased aperture area and efficiency feeding high and
ultrahigh--resolution spectrographs will, in coming years, provide accurate
enough spectra that tidally locked rotation of HD209458b has a detectable
influence.

\section{A Model of a Planetary Transit}
\label{sec:model}
We consider the spectrum of a star whose companion planet transits across the
face of the stellar disk from Earth's perspective.  The primary effect of the
planet is to reduce the stellar flux at all wavelengths, but the planet's
chemical composition, internal structure, and rotation rate influence
the spectrum in wavelength--dependent ways.  Since each
of these factors -- and others too, such as the star's rotation --
influences the observed spectrum, we built a model that incorporates
the many parameters related to each process.
The star and the planet are both assumed to rotate as solid bodies, with no
other (nonthermal) motion in their atmospheres.  Since deviations from
pure solid body rotation are likely to be no more than 25\% over the disk of
the star -- e.g., the Sun's equator--to--pole variation in rotation rate
is about 21\%, as per \citet{howard_et_al1984}, this is probably a reasonable
assumption for the star.  For the planet, this assumption might fail,
because wind--speeds in excess of the equatorial rotation speed of
$v_{\rm eq} \approx 2 {\rm ~km~s^{-1}}$ are predicted by many models, as
described in \S~\ref{sec:intro} above.  Still, when making this initial study
of the spectral effect of the motion of a transiting planet's atmosphere,
separating rotation from other processes makes the problem more tractable.
We set parameter values to match measured values from the HD209458b system
where possible.

The planet is modeled as an inner component that is entirely opaque and an
outer component that is isothermal and drops off exponentially. We compute
the wavelength--dependent optical depth due to the sodium D-doublet at
$\approx 590$~nm in the planet's atmosphere;
important parameters include the temperature and density of the planet's
atmosphere and its Na--content.  We use the Voigt profile -- as described
by, e.g., \citet{press+rybicki1993} -- to calculate $\tau[\lambda]$, the
optical depth to absorption along the line of sight.

As the planet transits the star, there are four points of
``contact'' between the planet and the star (really between their projections
on the sky): when the disk of the planet first touches the disk of the
star; when the planet is first entirely over the stellar disk; when the planet
is last entirely over the stellar disk; and when the planet last
touches the stellar disk.  We will additionally sometimes refer to
``$1.5^{\rm th}$'' contact (half--way between first and second contact),
and analogously to ``$2.5^{\rm th}$'' and ``$3.5^{\rm th}$'' contact.

As described in \S~\ref{sec:overview} above, the type of distortion that a
planet's rotation imposes relative to a non--rotating planet changes
depending on when during the transit the observation is made.  During
ingress or egress, the rotation of a tidally locked planet's atmosphere will
impose a distortion similar to an overall shift relative to a non--rotating planet: redshift during ingress; blueshift during egress.
When the planet is in mid--transit, in the middle of the stellar disk,
the overall distortion to the spectrum imposed by its rotation is
akin to a star's rotational broadening.  Since the line--centers of the lines
we are considering are heavily saturated and therefore flat at their
cores, rotational broadening has the somewhat counterintuitive effect
of steepening the cores of the profiles while broadening the wings.
We will discuss this in greater detail in the next section.
Although the type of distortion is different during ingress and egress from
during mid--transit, it turns out that the amount of distortion, in terms of
S/N ratio, is nearly constant throughout transit.  This, too, we will discuss
in \S~\ref{sec:results} below.

We simulate the HD209458b system, with a $1.32 R_J$ planet in a 3.5 day orbit,
orbiting a G0 star at with radius $1.05 R_\sun$ that is 47~pc away.  Our
model star has the limb darkening profile that \citet{knutson_et_al2007a}
measured for HD209458.  In order to approximate the fits to the data in
\citet{charbonneau_et_al2002}, we assign our model planet's atmosphere a
sodium--content and cloud deck height (1\% solar, and 0.01~bars) that are
comparable to the parameter--combinations that result in the best fits in
that paper. Finally, we present results at our simulation's spectral resolution
($R_S = 700,000$), and we simulate transit events observed using two different
lower resolution spectrographs, one with spectral resolution
$R^\prime_S = 50,000$ and one with $R^\prime_S = 150,000$.
All spectrographs (and associated optical paths) in our simulations have
100\% throughput efficiency.  In the remainder of this section, we provide
a detailed description of our parameterization of the problem.

\subsection{Parameters of the Star}
\label{ssec:pstar}
The parameters related to the star are listed in Table~\ref{ta:pstar}.
They are set to match measured parameters for HD209458, and we use the
limb--darkening profile from \citet{knutson_et_al2007a}.  We normalize
the flux to that of a blackbody of temperature $T_*$ of the size and at
the distance of HD209458.

\begin{deluxetable}{llr}
\tablewidth{400pt}
\tablecolumns{3}
\small
\tablecaption{Model Transit Parameters: Star}
\tablehead{\colhead{Parameter} & \colhead{Description} & \colhead{Value}}
\startdata
$M_*$         & Star Mass                      & 1.05 $M_\sun = 2.09\times 10^{33} {\rm~g}$\\
$R_*$         & Star Radius                    & 1.05 $R_\sun = 7.35\times 10^{10} {\rm~cm}$\\
$T_*$         & Star Temperature               & 6100 K\\
$d_*$         & Distance to star               & 47 pc\\
$\tau_*$      & Stellar Rotation Period        & 1 month\\
\enddata
\label{ta:pstar}
\vspace{-0.4cm}
\end{deluxetable}

\subsection{Parameters of the Planet}
\label{ssec:pplanet}
The parameters related to the planet are in Table~\ref{ta:pplanet}.
We model the planet as an inner component that is essentially a billiard ball
(completely opaque at all wavelengths) and an outer component that is an
isothermal atmosphere with scale height $H = R_{\rm gas}T_p/\mu g$, where
$R_{\rm gas}$ is the gas constant, $\mu$ is the molar mass, and $g$ is the
acceleration of gravity.  The density of our model planet's atmosphere varies
as $\rho = \rho_0 \exp[(r-{R_p}_0)/H]$, where ${R_p}_0$ is the radius of
the optically thick part (some authors have called this radius the
``cloud--deck'' \citep{charbonneau_et_al2002}).  This hypothetical
cloud deck could cause the planet to be optically thick at a higher
altitude than would otherwise be expected, as discussed in,
e.g., \citet{richardson_et_al2003} and \citet{sudarsky_et_al2000}.  The cloud
deck causes the optical depth as a function of radius in our model to have
a singular discontinuity at radius  ${R_p}_0$.

\begin{deluxetable}{llr}
\tablecolumns{3}
\small
\tablecaption{Model Transit Parameters: Planet}
\tablehead{\colhead{Parameter} & \colhead{Description} & \colhead{Value}}
\startdata
$M_p$         & Planet Mass                    & $0.69 M_J = 1.31\times 10^{30} {\rm~g}$\\
${R_p}_0$     & Optically Thick Planet Radius  & $1.32 R_J = 9.44\times 10^5 {\rm~km}$\\
$P_0$         & Planet Pressure at ${R_p}_0$   & 0.01 bars\\
$H$           & Planet Atmosphere Scale Height & 500 km\\
$T_p$         & Planet Atmosphere Temperature  & 1300 K\\
$f_{\rm TL}$  & Frac. Tidal Locked Rot. Rate   & \hspace{1in} 0 or 1 ($v_{\rm eq} = 0$ or $2 {\rm~km~s^{-1}}$)\\
$a$           & Semi--Major Axis               & 0.046 AU\\
$\#_H$         & Number of Scale Heights in Atm. & 15\\
\enddata
\label{ta:pplanet}
\vspace{-0.4cm}
\tablenotetext{a}{Parameter values are set to match measured values from
the HD209458b system where possible.
}
\end{deluxetable}

\subsection{Spectral Parameters}
\label{ssec:pspec}
The parameters pertaining to the shape of the observed spectrum are in
Table~\ref{ta:pspec}.  In addition to the location of the planet within the
stellar disk, the shape of the stellar spectrum and the wavelength--dependent
opacity of the planet's atmosphere together influence the transmission
spectrum.

{\tt Spec\_Shape} is a parameter that can take on the values
``Flat'', ``Blackbody'', or ``Solar'', and determines the rest--frame spectrum
of the model stellar photosphere.  (The integrated stellar spectrum is the
convolution of the rest--frame spectrum with the stellar rotation profile.)
When ``Flat'' is chosen, the rest--frame model stellar spectrum intensity is
set to the mean value of the blackbody intensity in the specified wavelength
range $[\lambda_{\rm min},\lambda_{\rm max}]$, which, in our simulation, is
set to $[580 {\rm~nm}, 600 {\rm~nm}]$.  When ``Solar'' is chosen, the
model stellar spectrum instensity is set to a high--resolution solar spectrum
that is normalized to the flux from
HD209458\footnote{From {\it ftp://solarch.tuc.noao.edu/}.}; but the Na D
lines in this high--resolution spectrum have been replaced by
Gaussian fits to the solar lines.

The planet's atmosphere has $N_{\rm abs}$ absorption features, each of which
is due to an element with a given fraction of the solar abundance.  In
the models presented in this paper, $N_{\rm abs} = 2$: we consider the Na
doublet at 588.9950 nm and 589.5924 nm, with sodium at fractional abundance
$f_\sun \equiv {X_{\rm Na~}}_p / {X_{\rm Na~}}_\sun  = 0.01$ of
the solar abundance.  Each line is modeled as a Voigt profile, as described
in, e.g., Press \& Rybicki (1993).

\begin{deluxetable}{llr}
\tablecolumns{3}
\small
\tablecaption{Model Transit Parameters: Spectral Features}
\tablehead{\colhead{Parameter} & \colhead{Description} & \colhead{Value}}
\startdata
{\tt Spec\_Shape} &  Shape of Star Spectrum    & Flat, Blackbody, or Solar\\
$\lambda_{\rm min}$ & Min. Wavelength in Sim.  & 580~nm\\
$\lambda_{\rm max}$ & Min. Wavelength in Sim.  & 600~nm\\
$N_{\rm abs}$ & \# Abs. Features in P. Atm.    & 2\\
${f_\sun}_1$  & Frac. Solar Abund., First Line & $0.01$\\
${\lambda_0}_1$ & First Line--Center           & 588.9950 nm\\
${A_{ki}}_1$  & Transition Prob. First Line    & $6.16\times 10^{-9} {\rm~s^{-1}}$\\
${g_i}_1$     & Stat. Wt. Lower Level First Line & 2\\
${g_k}_1$     & Stat. Wt. Upper Level First Line & 4\\
${f_\sun}_2$  & Frac. Solar Abund., Second Line & $0.01$\\
${\lambda_0}_2$ & Second Line--Center          & 589.5924 nm\\
${A_{ki}}_2$  & Transition Prob. Second Line    & $6.14\times 10^{-9} {\rm~s^{-1}}$\\
${g_i}_2$     & Stat. Wt. Lower Level Second Line & 2\\
${g_k}_2$     & Stat. Wt. Upper Level Second Line & 2\\
\enddata
\label{ta:pspec}
\vspace{-0.4cm}
\tablenotetext{a}{In parameters that have $i$ and $k$ subscripts,
$i$ indicates the lower level ($3 s_{1/2}$ for both lines) and $k$
indicates the upper level ($3 p_{3/2}$ for the bluer line and
$3 p_{1/2}$ for the redder line).  The fractional solar abundance is set
to 0.01 in order to achieve modest agreement with data observed for the Na D
doublet in HD209458b's atmosphere.
}
\end{deluxetable}

\subsection{Parameters of Observing and Computing}
\label{ssec:pobscomp}
The final set of parameters, listed in Table~\ref{ta:pobscomp}, includes
those that specify the observer and those that determine how the
observation is discretized for the purpose of numerical computation.
The model observational setup is determined by three parameters:
the telescope's diameter $D$ (6~m in our simulations) and efficiency
$\eta$ (100\%), and the spectrograph's spectral resolution $R_S$ (we set
$R_S$ to 700,000 for the purpose of computing the model, and we
re--bin to lower, more easily achieved resolutions -- 150,000 and
50,000 -- after computing a model).  These three parameters prescribe
the sizes of the spectral bins and the rate at which those bins
are capable of collecting light.

In order to compute the flux at Earth as a function of wavelength, we
begin by dividing the stellar disk into ${n_b}_*$ concentric annuli,
and we divide each annulus into ${n_\phi}_*$ azimuthal sections.  In
each section, the redshifted spectrum and the normalization must
both be computed.

Knowing the stellar rotation rate and axis, we may calculate the recessional
velocity of any point on the star's surface as a function of its location on
the projected stellar disk, and we redshift the spectrum from each part
of the star accordingly.
When the planet is in--transit, we separate the stellar disk into an annulus
that contains the planet and the rest of the disk that we treat as described
above.  The annulus that contains the planet is treated almost as above --
divided into ${n_b}_{*A}$ sub--annuli, each of which has ${n_\phi}_{*A}$
azimuthal sections -- but the sub--annuli are incomplete, interrupted by
the planet.

In order to sample the planet's atmosphere, we divide the region that overlaps
the star into ${n_b}_p$ concentric annuli around the planet's center,
each of which is divided into ${n_\phi}_p$ azimuthal sections.  In each
section, we must determine the optical depth and multiply by
$\exp{\left(-\tau\right)}$.  In calculating the optical depth, we note
that in the case that the planet's rotation axis is entirely normal to the
line--of--sight, if the planet rotates as a solid body then the radial
component of its recessional velocity is constant along a ray:
\begin{equation}
\tau[b_p, \phi, \lambda] = N[b_p] \times \sigma\left[\frac{\lambda}{1 + (v_p[b_p, \phi_p] /c)}\right],
\label{eq:tau}
\end{equation}
where the column density is calculated in terms of a function $G$ that
is specified below: $N[b_p] = n_0 G[b_p, {R_p}_0, H]$.  In
equation~(\ref{eq:tau}), $v_p[b_p, \phi_p]$ is the recessional velocity of
the planet, as a function of radius and azimuth, which depends upon
the orbit and the rotation.  Note that there is a single $v_p$ along a given
line--of--sight defined by a $(b_p,\phi_p)$ pair only under the assumption of
solid body rotation.  The rest--frame cross--section $\sigma[\lambda]$ is
computed according to the Voigt profile.  The function $G$ is defined as
the following
integral:
\begin{equation}
G[b_p, {R_p}_0, H] = \left\{ \begin{array}{ll}
\int_{-\infty}^{\infty} \exp\left[-\frac{\sqrt{b^2 + l^2} - {R_p}_0}{H}\right] d l & b_p > R_p \\
\infty & b_p \leq R_p
\end{array} \right. .
\label{eq:funcG}
\end{equation}

\begin{deluxetable}{llr}
\tablecolumns{3}
\small
\tablecaption{Model Transit Parameters: Observation and Computing}
\tablehead{\colhead{Parameter} & \colhead{Description} & \colhead{Value}}
\startdata
$D$           & Telescope Diameter             & $2.4 {\rm~m} - 30 {\rm~m}$\\
$\eta$        & Spectroscope Efficiency        & $1.00$ \\
$R^\prime_S$        & Obs. Spec. Resolution          & 50,000 - 700,000\\
$\mathcal{T}_{\rm int}$ & Integration Time     & $932.088 {\rm~s}$\\
--------------&--------------------------------&---------------------------\\
$R_S$         & Comp. Spec. Resolution         & 700,000\\
$\Delta t$    & Time--Step in Integration      & $50 {\rm~s}$\\
${n_b}_*$     & \# of Star Annuli               & 10\\
${n_\phi}_*$  & \# of Star Azimuthal Sections   & 16\\
${n_b}_{*A}$  & \# of S. Annuli in P. Annulus   & 10\\
${n_\phi}_{*A}$& \# of S. Azim. Sec.'s in P. Ann. & 10\\
${n_b}_p$     & \# of Planet Atm. Annuli        & 20\\
${n_\phi}_p$  & \# of Planet Atm. Azim. Sections & 20\\
\enddata
\label{ta:pobscomp}
\vspace{-0.4cm}
\tablenotetext{a}{Parameter values are set to match measured values from
the HD209458b system where possible.
}
\end{deluxetable}

\section{Model Transit Spectra}
\label{sec:results}
As described in \S~\ref{ssec:preview}, we seek the expected signal--to--noise
ratio for distinguishing between the spectrum that would be observed due
to a non--rotating planet (or one that is is in a Cassini state with its
rotation axis nearly in the plane of orbit) and the spectrum that would be
observed due to a tidally locked planet.
The computed model spectrum $\mathcal{N}[\lambda]$ is the time integral
of the instantaneous spectrum $\dot{\mathcal{N}}[\lambda]$ and consists
of the number of photons detected per wavelength bin:
\[
\mathcal{N}[\lambda] \approx \dot{\mathcal{N}}[\lambda] \Delta t_{\rm obs},
\]
for some small exposure time $\Delta t_{\rm obs}$.

The model signal (of rotation) per bin that we are looking for is the
difference between the rotating model spectrum $\mathcal{N}_{\rm rot}$
and the non--rotating model spectrum $\mathcal{N}_{\rm no~rot}$:
\begin{equation}
S_b = \left(\dot{\mathcal{N}}_{\rm rot}[\lambda] - \dot{\mathcal{N}}_{\rm no~rot}[\lambda]\right) \Delta t_{\rm obs}.
\label{eq:signal}
\end{equation}
We make the optimistic approximation that the noise per bin is just the
photon--noise:
\begin{equation}
N_b = \sqrt{\dot{\mathcal{N}}_{\rm no~rot}[\lambda] \Delta t_{\rm obs}}.
\label{eq:noise}
\end{equation}
The total signal--to--noise ratio in a single exposure, then, is the sum in
quadrature of $S_b/N_b$ for all wavelength bins $\lambda_i$:
\begin{equation}
S/N = \sqrt{\sum_{i = 1}^{\# \rm bins} \left( \frac{\dot{\mathcal{N}}_{\rm rot}[\lambda_i] - \dot{\mathcal{N}}_{\rm no~rot}[\lambda_i]}{\sqrt{\dot{\mathcal{N}}_{\rm no~rot}[\lambda_i]}} \right)^2 } \times \sqrt{\Delta t_{\rm obs}}.
\label{eq:sn}
\end{equation}
A similar summation in quadrature applies over all exposures.
Note that, in principle, the expression in equation~(\ref{eq:sn}) is
insensitive to the sizes of bins and hence to the spectral resolution $R_S$,
as long as the bins are small relative to the Gaussian width of the
absorption feature under consideration.  Our simulations indicate that the
spectral resolution must be $\gsim 500,000$ in order for S/N to be nearly
independent of $R_S$.

\begin{figure}[h!]
\plotone
{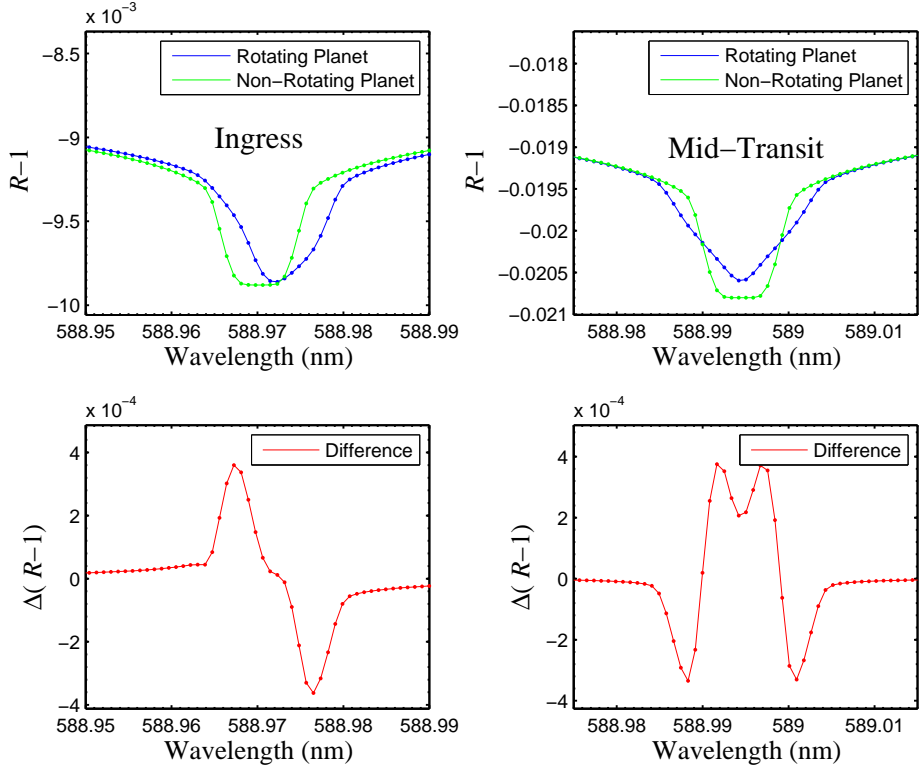}
\caption{Upper panels show snap--shot spectra for one of the the Na D lines
for two different model planets (tidally locked and non--rotating); lower
panels show the difference between the two model spectra.The quantities
plotted are $\mathcal{R}' = \mathcal{R}-1$ (upper panels) and
$\Delta \mathcal{R}'$ (lower panels), where
$\mathcal{R}[\lambda,t] = \dot{N}_{\rm in~transit}[\lambda,t]/\dot{N}_{\rm out~of~transit}[\lambda,t]$.
In the upper panels, the blue curve is the tidally locked planet's transit
spectrum, and the green curve is the non--rotating planet's transit spectrum.
In the lower panels, the difference between the rotating and non--rotating
planet's spectra.
{\it Left}: Halfway through ingress (at $1.5^{\rm th}$ contact).
{\it Right}: Halfway through the whole transit ($2.5^{\rm th}$ contact).}
\label{fig:spec. and diff.}
\end{figure}

The effect of rotation, both during ingress and during mid--transit,
is illustrated in Figure~\ref{fig:spec. and diff.}.
For illustrative purposes, in this figure we assume a uniform star (flat
spectrum, non--rotating, no limb--darkening).  On the left panels of
Figure~\ref{fig:spec. and diff.}, we show a snap--shot during ingress
(at $1.5^{\rm th}$ contact) and on the right, we show a snap--shot during the
middle of a transit ($2.5^{\rm th}$ contact).  The quantity plotted is
$\mathcal{R}' = \mathcal{R}-1$ from B01, where
\begin{equation}
\mathcal{R}[\lambda,t] = \frac{\dot{N}_{\rm in~transit}[\lambda,t]}{\dot{N}_{\rm out~of~transit}[\lambda,t]}
\label{eq:Brown R}
\end{equation}
The bottom panels of Figure~\ref{fig:spec. and diff.} show the
difference spectra between the models with a tidally locked planet and the
models with a non--rotating planet ($\mathcal{R}_{\rm rot} - \mathcal{R}_{\rm no~rot}$).

\begin{figure}[h!]
\plotone
{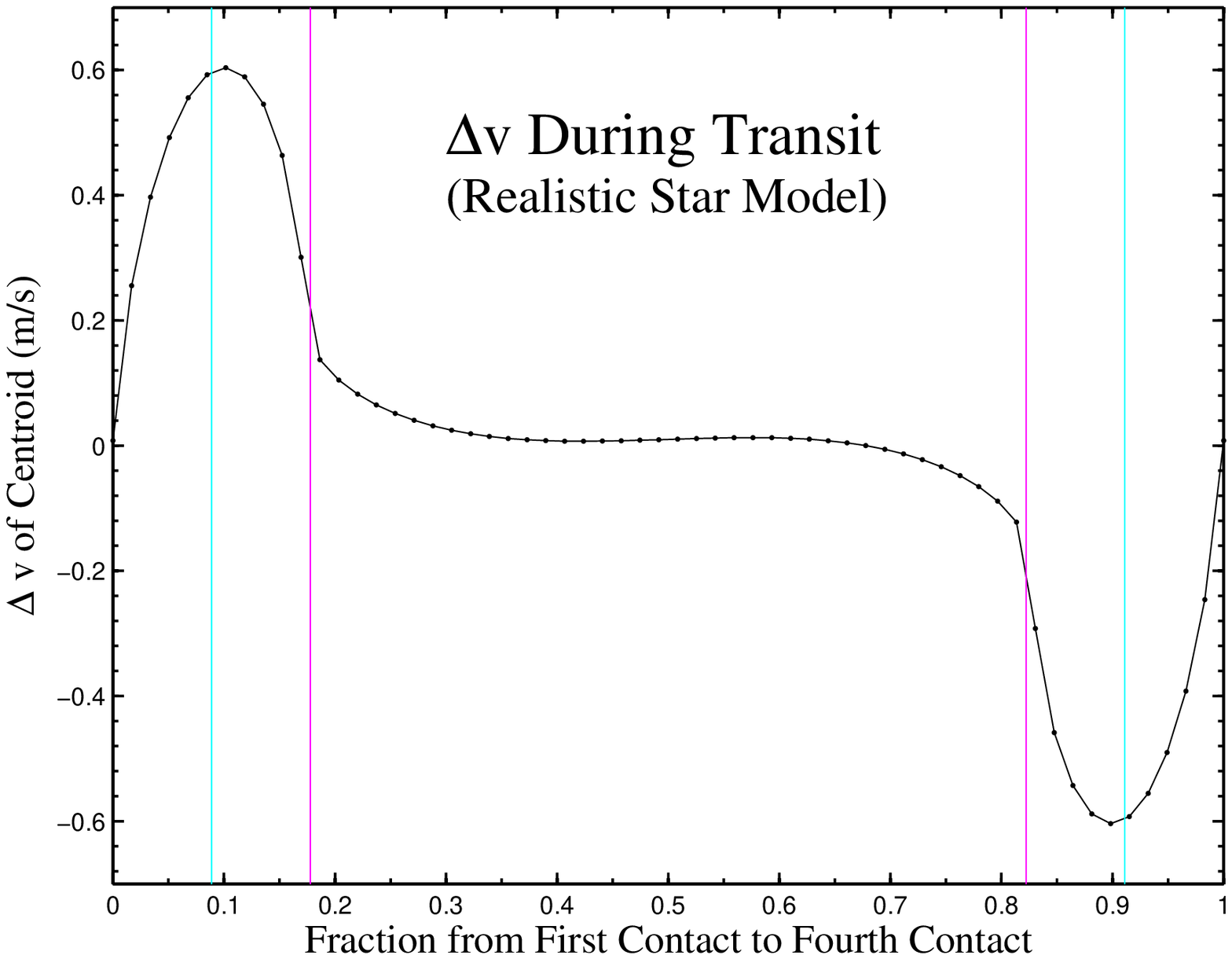}
\caption{Centroid--shift of Na D lines from tidally locked rotation from the
beginning to the end of a transit, relative to an identical but non--rotating
planet; Sun--like stellar spectrum.
The vertical lines denote $1.5^{\rm th}$ and $3.5^{\rm th}$ contact (cyan)
and second and third contact (magenta).
Between first and second contact, the
spectrum with the rotating planet is redshifted relative to the non--rotating
planet by up to about $60 {\rm ~cm~s^{-1}}$; between third and fourth contact,
it is blueshifted by the same amount.  This plot samples the transit at 60
regularly--spaced points.  Parameters were chosen to represent the HD209458
system.}
\label{fig:delta v}
\end{figure}

As described in \S~\ref{sec:overview} above, a planet's rotation causes
the centroids of stellar absorption features to shift relative to a
non--rotating planet.  In Figure~\ref{fig:delta v}, centroid shifts
(in velocity units) are plotted as a function of position in transit, for
a planet transiting in front of a realistic star model with a Sun--like
spectrum. The recessional velocity increases roughly sinusoidally during
ingress, reaching a peak of about $60 {\rm ~cm~s^{-1}}$ at $1.5^{\rm th}$
contact.  During mid--transit, between $2^{\rm nd}$ and $3^{\rm rd}$
contacts, the net velocity shift is much smaller.  Egress is nearly
perfectly symmetrical with ingress, though the velocity shifts have the
opposite sign.

\begin{figure}[h!]
\plotone
{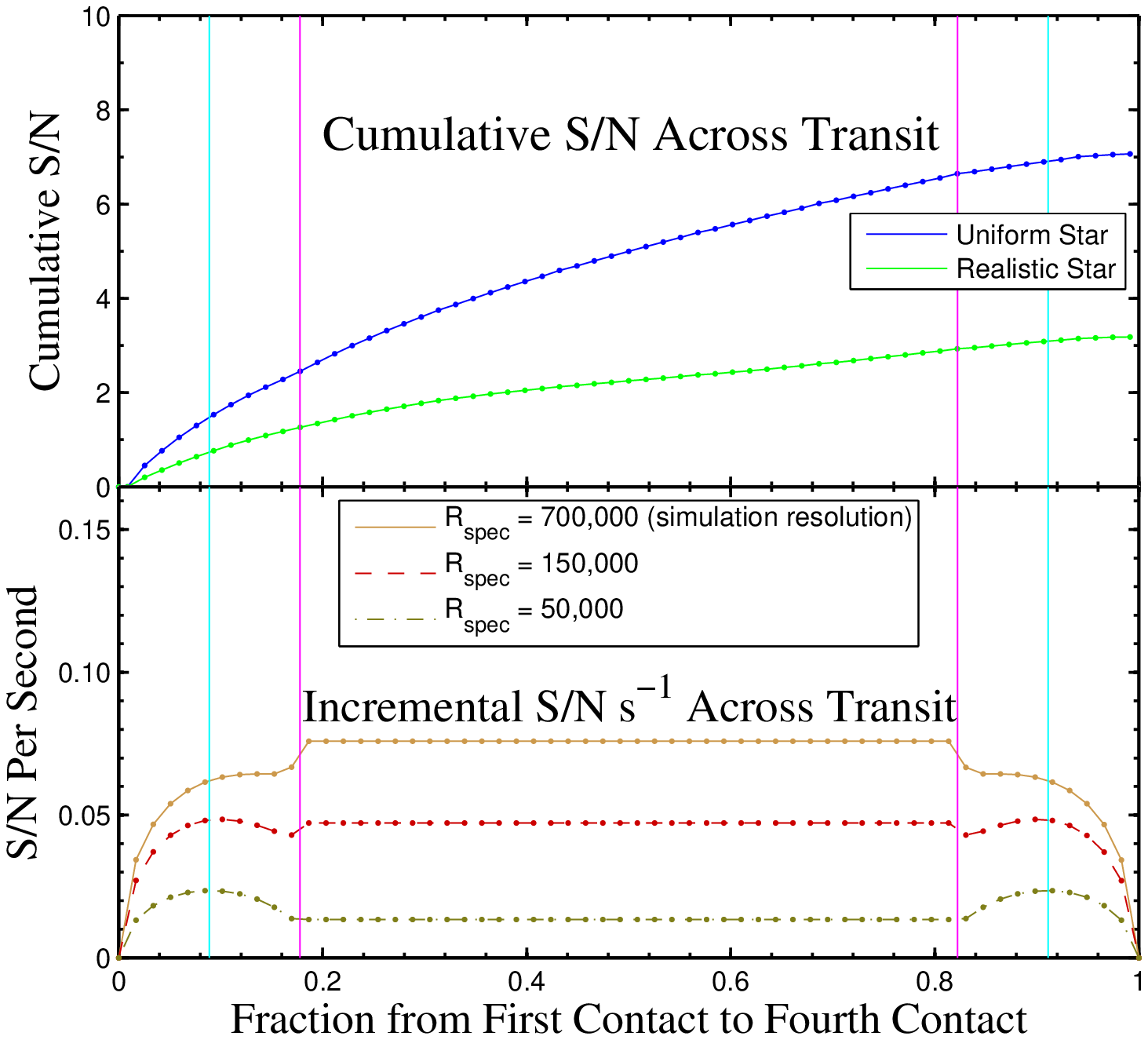}
\caption{{\it Top}: S/N for distinguishing between a rotating and a
non--rotating planet, accumulated during a single transit.  Upper curve
displays the cumulative S/N for the distinction of a tidally locked planet
from a non-rotating model, assuming a completely uniform star: the model star
has a flat spectrum, is not rotating, and there is no limb--darkening.
Lower curve displays the same, but for a more realistic star: the model
star rotates once per month, has the limb--darkening profile determined by
\citet{knutson_et_al2007a}, and has a solar spectrum.
These curves assume the spectral resolution used in the simulation
($R_S = 700,000$); for more easily achievable spectral resolution, these
curves are adjusted downward by 30\% - 80\%, as displayed in the bottom panel.
The vertical lines denote $1.5^{\rm th}$ and $3.5^{\rm th}$ contact (cyan)
and second and third contact (magenta).
Also, both top and bottom panels show model results assuming 100\% throughput
efficiency ($\eta = 100\%$).  In practice, these curves would
be adjusted downward by a factor $\eta^{1/2}$, or $\sim 0.04^{1/2} = 0.2$ for
an ultrahigh resolution spectrograph.
{\it Bottom}: Incremental S/N for sixty 1--second observations across the
transit, assuming a uniform, non--rotating star with a flat spectrum.  Top
curve shows the incremental S/N, assuming a spectrograph with the spectral
resolution of the simulation ($R_S = 700,000$).  Middle and bottom curves
show the same, assuming spectral resolutions $R^\prime_S = 150,000$ and
$R^\prime_S = 50,000$, respectively.  Parameters of the system (except for
the stellar spectrum in the bottom panel) were chosen to represent the
HD209458 system.}
\label{fig:inc. and cum.}
\end{figure}

The cumulative and incremental signal--to--noise ratio across the transit
are shown in the top and bottom panels, respectively, of
Figure~\ref{fig:inc. and cum.}.  As a planet proceeds in its transit from
first contact to fourth contact, the S/N builds up steadily throughout
the transit, as can be seen in the top panel of
this figure.  This cumulative increase reflects the steady
incremental S/N per second of observation, which, for a planet
crossing in front of a uniform star, is shown in the bottom panel.
The three curves in the bottom panel represent, in decreasing
order of S/N, the incremental S/N curves that are expected for a spectrograph
with ultrahigh--resolution (the simulation's resolution $R_S = 700,000$),
for a spectrograph with as high--resolution as bHROS on Gemini
($R^\prime_S = 150,000$), and for a spectrograph with more standard
high--resolution
($R^\prime_S = 50,000$).  In the top panel, we show the cumulative S/N (assuming
the simulation's resolution) for a planet in front of a uniform star, and for
a planet in front of a more realistic sun--like star that rotates once per
month, has a limb--darkening profile, and has a solar spectrum.  It
is apparent in the top panel of Figure~\ref{fig:inc. and cum.} that a
spectrograph with our simulation's resolution would see the effect of rotation
at $S/N = 7.1 \eta^{1/2}$ in a single transit in the case of the simplified
uniform star, and at $S/N = 3.2 \eta^{1/2}$ in one transit of the realistic star.
The effect of including in the simulation the realistic features of stellar rotation,
limb--darkening, and a solar spectrum is therefore to depress the S/N of the
effect of tidally locked rotation by a factor of slightly more than 2.  The bottom
panel of the figure indicates that our predicted S/N for one transit of a realistic
star ($3.2 \eta^{1/2}$) might be adjusted downward by $\sim 50\%$, depending on the
spectral resolution, indicating a total one--transit S/N for the case of the
realistic star of about $\sim 1.7 \eta^{1/2}$.  Finally, we do not show the
incremental S/N for the realistic star model, but, as the top panel of the figure
indicates, a larger proportion of the total signal comes from ingress and egress
in the realistic star model than in the uniform star model.

\section{Discussion}
\label{sec:disc}
The $S/N$ scales with the square root of the number of absorption lines under
consideration, with the square root of the number of transits, and with
several other parameters, as follows:
\begin{eqnarray}
\nonumber S/N & \sim & 
  \left( \sqrt{\frac{N_{\rm abs}}{2} } \right)
  \left( \sqrt{ \#~\rm transits} \right)
  \left( \frac{D \sqrt{\eta}}{\rm 6~m} \right) \times \\
\nonumber & &
  10^{-0.2(V_*-V_0)} \times
  \left(\frac{R_*}{R_0}\right)^{-2}
  \left(\frac{R_p H}{{R_p}_0 H_0}\right)
  \left(\frac{v}{2000 \rm~m~s^{-1}}\right) \times \\
\label{eq:snscale} & &
  \left(\psi\left[R^\prime_S\right]\right)
  \left( S/N \right)_0
\end{eqnarray}
In this equation, $N_{\rm abs}$ is the number of absorption lines,
$D$ is the telescope's diameter, $\eta$ is the efficiency of the instrumental
setup, $V_*$ is the star's magnitude, $V_0$ is HD209458's magnitude
(about 7.6), $R_*$ is the stellar radius, $R_0$ is HD209458's radius,
$R_p$ and $H$ are the planet's radius and scale height, and ${R_p}_0$ and
$H_0$ are HD209458b's radius and scale height.
$\psi[R^\prime_S]$ is a function that characterizes the (nonlinear)
response of S/N to the spectral resolution of the spectrograph; for instance,
$\psi[700,000] \approx 1$, $\psi[150,000] \sim 0.6$ and
$\psi[50,000] \sim 0.2$.  Finally, $(S/N)_0 = 7.1$ for a uniform star and
${X_{\rm Na~}}_p/{X_{\rm Na~}}_\sun = 0.01$; and $(S/N)_0 = 3.2$ for a more
realistic (rotating, limb--darkened) star with a solar spectrum.

Culling the signal from mid--transit, however, is much more difficult
than from ingress and egress, because the shape of the distortion depends
more sensitively on the structure of the planet.  For the realistic
star, our model predicts $S/N \sim 1.3 \eta^{1/2}$ (with the 6~m
telescope, with Poisson--dominated photon noise) for ingress alone, and
therefore a factor $2^{1/2}$ greater, or $S/N \sim 1.8 \eta^{1/2}$ for
ingress and egress.

In Table~\ref{ta:norbits}, we present the number of orbits that must be
observed in order to make a $5\sigma$ detection of rotation, for various
combinations of parameters.  In all cases, we assume a realistically achievable
optical setup, with spectral resolution $R^\prime_S = 150,000$ and
throughput $\eta = 4\%$.  Improvement in $R^\prime_S$ will yield a modest
improvement in S/N, and improvement in $\eta$ could be quite significant.
In the third column of Table~\ref{ta:norbits} (A), we present the total
available S/N for the whole transit, according to our model, while in the
fourth column (B), we present the available S/N from just ingress and egress.
The fifth and sixth columns are based on the S/N in the fourth column.
\begin{deluxetable}{crcccr}
\tablecolumns{6}
\small
\tablecaption{Required Number of Transits For $5\sigma$ Detection}
\tablehead{\colhead{{\tt Spec\_Shape}}  & \colhead{$D$} & \colhead{(S/N)/Tr. (A)} & \colhead{(S/N)/Tr. (B)} & \colhead{Req. \# Transits (A)} & \colhead{Duration (A)}}
\startdata
Flat (uniform star)  &  6~m  &  1.1  &  0.52 &  $\sim 25$   & $\sim 3$  months \\
Solar                &  6~m  &  0.48 &  0.27 &  $\sim 110$  & $\sim 1.25$ years \\
Solar                &  10~m &  0.79 &  0.45 &  $\sim 40$   & $\sim 5$ months \\
Solar                &  30~m &  2.4  &  1.3  &  $\sim 5$    & $\sim 3$  weeks \\
\enddata
\label{ta:norbits}
\vspace{-0.4cm}
\tablenotetext{a}{A table showing the number of orbits required to achieve a
$S/N =5$ detection of tidally locked rotation of HD209458b, assuming a
planetary Na--content that is 1\% solar.  Assumptions: we are monitoring two
spectral lines (the Na doublet) with a spectrograph with resolving power
$R^\prime_S = 150,000$, and an efficiency in the optical setup of $0.04$.
The third column ``(S/N)/Tr. (A)'' is the S/N per transit, as shown in
Figure~\ref{fig:inc. and cum.}, and the fourth column is the S/N per
transit, considering only ingress and egress.  The ``(A)'' in the fifth and
sixth columns indicates that the numbers of transits and the required
duration are tabulated for S/N values from column three.
}
\end{deluxetable}

\begin{figure}[h!]
\plotone
{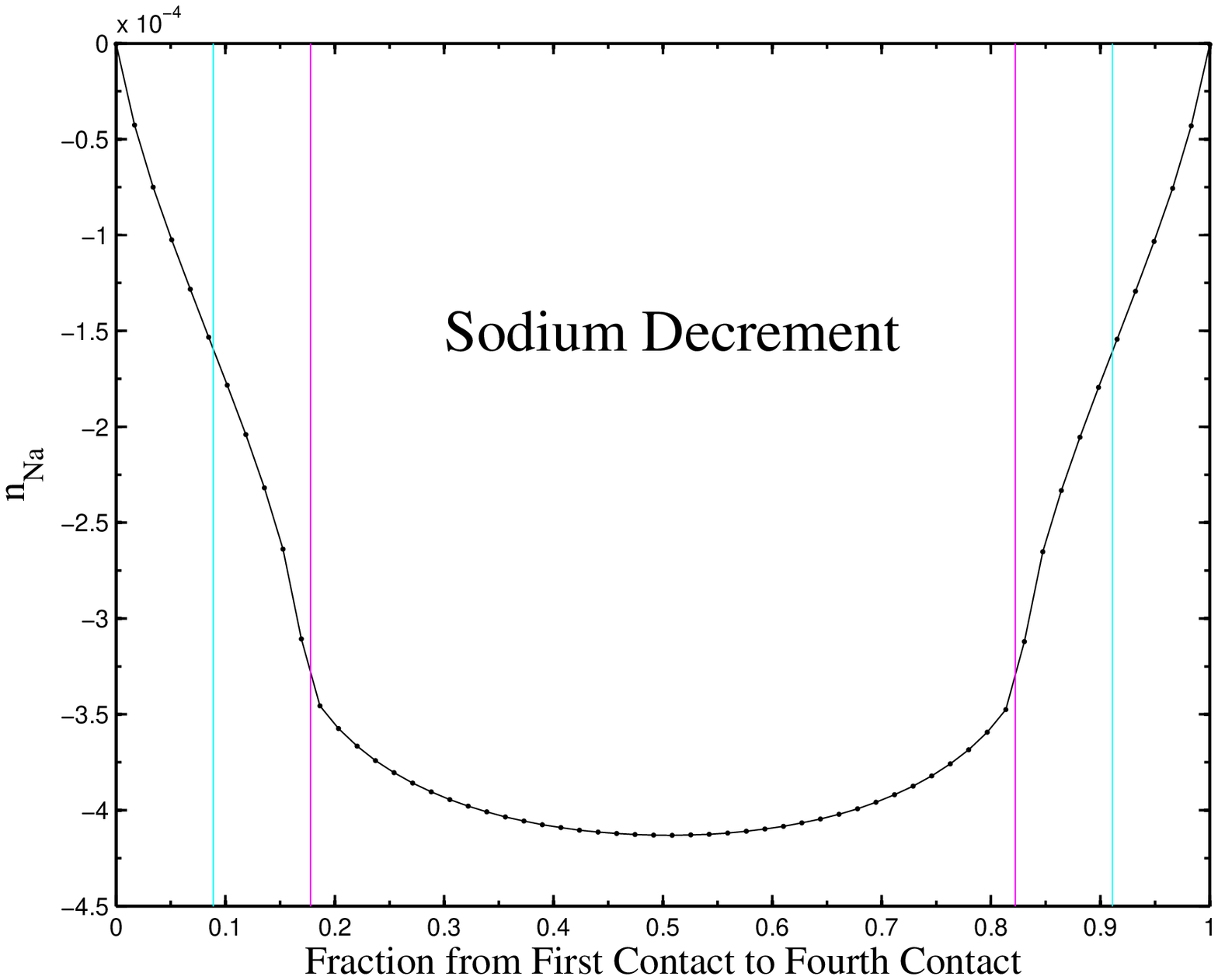}
\caption{The flux decrement due to sodium in the planet's atmosphere,
as shown by the quantity $n_{\rm Na}$ defined by \citet{charbonneau_et_al2002}.
This decrement reaches a maximum in magnitude of $-4.1 \times 10^{-4}$,
halfway through transit, in our model.}
\label{fig:charbonneau}
\end{figure}

Finally, as a sanity check, our model can be tested by comparing
it to the analysis of
\citet{charbonneau_et_al2002}.  That analysis suggests that
the sodium content of a planet's atmosphere can be determined
by comparing the flux in a narrow band centered on the sodium
resonance lines with the flux in a wider band surrounding but
excluding the lines.  In that paper, the decrement in the
narrow band containing the sodium features is named $n_{\rm Na}$,
and was measured to be $-2.32\times 10^{-4}$ for HD209458 during
the middle of transit.  They presented several models, all of
which over--predicted the sodium decrement.  The model that
predicted the smallest magnitude of the decrement had 1\% solar
metalicity and cloud deck at $0.0368 {\rm ~bar}$; this model
predicts $n_{\rm Na} \sim -3.4 \times 10^{-4}$ in mid--transit.
Our model (1\% solar metalicity, cloud deck at $0.01 {\rm ~bar}$)
predicts $\sim -4.1 \times 10^{-4}$ in mid--transit, as shown
in Figure~\ref{fig:charbonneau}.  We conclude that our model is in
reasonable agreement with both Charbonneau et al.'s model ($\sim 20\%$)
and with the actual data ($\sim 40\%$).

\section{Conclusion}
\label{sec:conc}
Our investigation indicates that, with currently available instruments,
it will be difficult to obtain the sensitivity needed to achieve a
minimal $S/N \gsim 5$ detection of tidally locked rotation of the
planet HD209458b.  Nevertheless, it appears that the effect of rotation
will have significant (though small) influence on transit--spectra taken with
current or near--future instruments.
Because this influence is so small, it is worth considering ways that the
$S/N$ could be improved, according to the scalings summarized in
equation~(\ref{eq:snscale}).

\begin{itemize}
\item The most obvious way to boost S/N is to increase the collecting area of
the telescope, the effect of which is shown in Table~\ref{ta:norbits}.  With an
extremely large (30~m--class) telescope, the required discrimination power
could be achieved in $\sim 5-15$ transits, if our optimistic assumptions about
the noise are not too far off.
\item As the abundances of other elements or molecules in the planet's
atmosphere are identified, the number of absorption lines $N_{\rm abs}$ that
can be utilized can increase.  Observing
four absorption features instead of two will boost the $S/N$ per orbit by
$2^{1/2}$ and will therefore cut in half the required observing time to
achieve a fixed target $S/N$.
\item The HD189733 system is only 19.3~pc away \citep{bouchy_et_al2005},
but the star is smaller and dimmer than HD209458, and has the same apparent
magnitude.  We find that, overall, it has no relative advantage in terms of
$S/N$.  It may be unlikely that a star significantly brighter than HD209458
will be found to have a transiting giant planet companion, but if this should
happen the signature of rotation will be more readily apparent in the system
with the brighter star.
\item Finally, since S/N is proportional to the size of the atmosphere,
represented by $\delta_{\rm atm}$ in equation~(\ref{eq:SN estimate}), a planet
with a more extended atmosphere would show the distortion due to rotation more
readily.  The size of the atmosphere depends on the product of $R_p$ and the
height of the atmosphere $\Pi_{\rm atm} \sim 15 H$, so the S/N could be
increased in the case of a larger planet, a hotter one, or one with lower
surface gravity.  Unfortunately, although some other planets are somewhat
larger, none of the puffier planets that have been discovered so far happen to
transit a star as bright as HD209458.  If a more bloated planet is found to
transit a star as bright as HD209458, such a planet would probably be a more
promising target for the type of study described in this paper.
\end{itemize}

 Although the optical transit spectrum of an extrasolar giant planet contains
many precious clues regarding the nature of rotation and climate or weather, we
emphasize that we have so far addressed only the forward problem: deep,
high spectral resolution transit spectra {\it will} be distorted by
HD209458b's tidally locked rotation, if indeed it is tidally locked.
But seeing the effect of rotation in a spectrum and {\it knowing} that we are
seeing the effect of rotation are two very different things.
The inverse problem -- identifying the rotation rate -- will be much more
challenging, and is beyond the scope of this paper's investigation.
In brief, it is impossible to decode the clues that the spectrum holds
without a Rosetta stone -- namely, without already possessing accurate
knowledge of abundances in the planet's atmosphere.
We note that when it comes to solving the inverse problem, it may be possible
to make do without a perfect Rosetta stone by comparing the star's spectrum
at two different times of transit instead of comparing it to an accurate
input model.  Specifically, comparing the ingress spectrum to the egress
spectrum should eliminate various systematic uncertainties that may effect both
ingress and egress spectra.  This strategy involves foregoing spectra between
second and third contacts, at a cost of a factor of $\sim 2$ in $S/N$, but has
the advantage of being more robust with respect to errors in modeling the
planetary atmosphere. Furthermore, if detailed analysis along the lines of
\citet{charbonneau_et_al2002} should reveal many more planetary absorption
features, a large current--generation telescope could detect the motion of an
extrasolar planet's atmosphere in of order a month.

A final point to keep in mind is that the technique presented in this paper
is not limited to discerning the difference between a tidally locked planet
and a non--rotating planet; this technique can be used to investigate the
applicability of any  model of the motion of a transiting planet's atmosphere.
Some atmospheric circulation models of Hot Jupiters predict extremely fast
winds, up to $\sim 9 {\rm ~km~s^{-1}}$, or more then four times the
predicted equatorial rotation rate.  It might therefore turn out to be easier
to detect winds than to detect rotation.  Moreover, if an analysis similar to
the one described here were to find a deviation from the hypothesis of
tidally locked rotation, this would mean that {\it something} interesting
is going on -- the planet could be rotating roughly as a solid body but at
a different rate; there could be significant non--rotational motion in
the atmosphere; or there could be some combination of these phenomena.

\vspace{\baselineskip}
\vspace{\baselineskip}
We thank Caleb Scharf, Frits Paerels, and Greg Bryan for extremely helpful
discussions.  We gratefully acknowledge George Rybicki for use of a Voigt
profile routine.  We acknowledge many helpful comments by our anonymous
referee.

\bibliography{biblio.bib}

\begin{thebibliography}{84}
\expandafter\ifx\csname natexlab\endcsname\relax\def\natexlab#1{#1}\fi

\bibitem[{{Alonso} {et~al.}(2004){Alonso}, {Brown}, {Torres}, {Latham},
  {Sozzetti}, {Mandushev}, {Belmonte}, {Charbonneau}, {Deeg}, {Dunham},
  {O'Donovan}, \& {Stefanik}}]{alonso_et_al2004}
{Alonso}, R., {Brown}, T.~M., {Torres}, G., {Latham}, D.~W., {Sozzetti}, A.,
  {Mandushev}, G., {Belmonte}, J.~A., {Charbonneau}, D., {Deeg}, H.~J.,
  {Dunham}, E.~W., {O'Donovan}, F.~T., \& {Stefanik}, R.~P. 2004, \apjl, 613,
  L153

\bibitem[{{Bakos} {et~al.}(2006){Bakos}, {Knutson}, {Pont}, {Moutou},
  {Charbonneau}, {Shporer}, {Bouchy}, {Everett}, {Hergenrother}, {Latham},
  {Mayor}, {Mazeh}, {Noyes}, {Queloz}, {P{\'a}l}, \& {Udry}}]{bakos_et_al2006}
{Bakos}, G.~{\'A}., {Knutson}, H., {Pont}, F., {Moutou}, C., {Charbonneau}, D.,
  {Shporer}, A., {Bouchy}, F., {Everett}, M., {Hergenrother}, C., {Latham},
  D.~W., {Mayor}, M., {Mazeh}, T., {Noyes}, R.~W., {Queloz}, D., {P{\'a}l}, A.,
  \& {Udry}, S. 2006, \apj, 650, 1160

\bibitem[{{Bakos} {et~al.}(2007{\natexlab{a}}){Bakos}, {Kovacs}, {Torres},
  {Fischer}, {Latham}, {Noyes}, {Sasselov}, {Mazeh}, {Shporer}, {Butler},
  {Stefanik}, {Fernandez}, {Sozzetti}, {Pal}, {Johnson}, {Marcy}, {Winn},
  {Sipocz}, {Lazar}, {Papp}, \& {Sari}}]{bakos_et_al2007bb}
{Bakos}, G.~A., {Kovacs}, G., {Torres}, G., {Fischer}, D.~A., {Latham}, D.~W.,
  {Noyes}, R.~W., {Sasselov}, D.~D., {Mazeh}, T., {Shporer}, A., {Butler},
  R.~P., {Stefanik}, R.~P., {Fernandez}, J.~M., {Sozzetti}, A., {Pal}, A.,
  {Johnson}, J., {Marcy}, G.~W., {Winn}, J., {Sipocz}, B., {Lazar}, J., {Papp},
  I., \& {Sari}, P. 2007{\natexlab{a}}, ArXiv e-prints, 705

\bibitem[{{Bakos} {et~al.}(2007{\natexlab{b}}){Bakos}, {Noyes}, {Kov{\'a}cs},
  {Latham}, {Sasselov}, {Torres}, {Fischer}, {Stefanik}, {Sato}, {Johnson},
  {P{\'a}l}, {Marcy}, {Butler}, {Esquerdo}, {Stanek}, {L{\'a}z{\'a}r}, {Papp},
  {S{\'a}ri}, \& {Sip{\H o}cz}}]{bakos_et_al2007a}
{Bakos}, G.~{\'A}., {Noyes}, R.~W., {Kov{\'a}cs}, G., {Latham}, D.~W.,
  {Sasselov}, D.~D., {Torres}, G., {Fischer}, D.~A., {Stefanik}, R.~P., {Sato},
  B., {Johnson}, J.~A., {P{\'a}l}, A., {Marcy}, G.~W., {Butler}, R.~P.,
  {Esquerdo}, G.~A., {Stanek}, K.~Z., {L{\'a}z{\'a}r}, J., {Papp}, I.,
  {S{\'a}ri}, P., \& {Sip{\H o}cz}, B. 2007{\natexlab{b}}, \apj, 656, 552

\bibitem[{{Barman}(2007)}]{barman2007}
{Barman}, T. 2007, \apjl, 661, L191

\bibitem[{{Barnes} \& {Fortney}(2003)}]{barnes+fortney2003}
{Barnes}, J.~W. \& {Fortney}, J.~J. 2003, \apj, 588, 545

\bibitem[{{Bodenheimer} {et~al.}(2003){Bodenheimer}, {Laughlin}, \&
  {Lin}}]{bodenheimer_et_al2003}
{Bodenheimer}, P., {Laughlin}, G., \& {Lin}, D.~N.~C. 2003, \apj, 592, 555

\bibitem[{{Bouchy} {et~al.}(2005{\natexlab{a}}){Bouchy}, {Bazot}, {Santos},
  {Vauclair}, \& {Sosnowska}}]{bouchy_et_al2005b}
{Bouchy}, F., {Bazot}, M., {Santos}, N.~C., {Vauclair}, S., \& {Sosnowska}, D.
  2005{\natexlab{a}}, \aap, 440, 609

\bibitem[{{Bouchy} {et~al.}(2004){Bouchy}, {Pont}, {Santos}, {Melo}, {Mayor},
  {Queloz}, \& {Udry}}]{bouchy_et_al2004}
{Bouchy}, F., {Pont}, F., {Santos}, N.~C., {Melo}, C., {Mayor}, M., {Queloz},
  D., \& {Udry}, S. 2004, \aap, 421, L13

\bibitem[{{Bouchy} {et~al.}(2005{\natexlab{b}}){Bouchy}, {Udry}, {Mayor},
  {Moutou}, {Pont}, {Iribarne}, {da Silva}, {Ilovaisky}, {Queloz}, {Santos},
  {S{\'e}gransan}, \& {Zucker}}]{bouchy_et_al2005}
{Bouchy}, F., {Udry}, S., {Mayor}, M., {Moutou}, C., {Pont}, F., {Iribarne},
  N., {da Silva}, R., {Ilovaisky}, S., {Queloz}, D., {Santos}, N.~C.,
  {S{\'e}gransan}, D., \& {Zucker}, S. 2005{\natexlab{b}}, \aap, 444, L15

\bibitem[{{Brown}(2001)}]{brown2001}
{Brown}, T.~M. 2001, \apj, 553, 1006

\bibitem[{{Burke} {et~al.}(2007){Burke}, {McCullough}, {Valenti},
  {Johns-Krull}, {Janes}, {Heasley}, {Summers}, {Stys}, {Bissinger}, {Fleenor},
  {Foote}, {Garcia-Melendo}, {Gary}, {Howell}, {Mallia}, {Masi}, {Taylor}, \&
  {Vanmunster}}]{burke_et_al2007}
{Burke}, C.~J., {McCullough}, P.~R., {Valenti}, J.~A., {Johns-Krull}, C.~M.,
  {Janes}, K.~A., {Heasley}, J.~N., {Summers}, F.~J., {Stys}, J.~E.,
  {Bissinger}, R., {Fleenor}, M.~L., {Foote}, C.~N., {Garcia-Melendo}, E.,
  {Gary}, B.~L., {Howell}, P.~J., {Mallia}, F., {Masi}, G., {Taylor}, B., \&
  {Vanmunster}, T. 2007, ArXiv e-prints, 705

\bibitem[{{Burrows} {et~al.}(2007){Burrows}, {Hubeny}, {Budaj}, \&
  {Hubbard}}]{burrows_et_al2007}
{Burrows}, A., {Hubeny}, I., {Budaj}, J., \& {Hubbard}, W.~B. 2007, \apj, 661,
  502

\bibitem[{{Burrows} {et~al.}(2004){Burrows}, {Hubeny}, {Hubbard}, {Sudarsky},
  \& {Fortney}}]{burrows_et_al2004}
{Burrows}, A., {Hubeny}, I., {Hubbard}, W.~B., {Sudarsky}, D., \& {Fortney},
  J.~J. 2004, \apjl, 610, L53

\bibitem[{{Burrows} {et~al.}(2003){Burrows}, {Sudarsky}, \&
  {Hubbard}}]{burrows_et_al2003}
{Burrows}, A., {Sudarsky}, D., \& {Hubbard}, W.~B. 2003, \apj, 594, 545

\bibitem[{{Butler} {et~al.}(1996){Butler}, {Marcy}, {Williams}, {McCarthy},
  {Dosanjh}, \& {Vogt}}]{butler_et_al1996}
{Butler}, R.~P., {Marcy}, G.~W., {Williams}, E., {McCarthy}, C., {Dosanjh}, P.,
  \& {Vogt}, S.~S. 1996, \pasp, 108, 500

\bibitem[{{Chabrier} \& {Baraffe}(2007)}]{chabrier+baraffe2007}
{Chabrier}, G. \& {Baraffe}, I. 2007, \apjl, 661, L81

\bibitem[{{Charbonneau} {et~al.}(2000){Charbonneau}, {Brown}, {Latham}, \&
  {Mayor}}]{charbonneau_et_al2000}
{Charbonneau}, D., {Brown}, T.~M., {Latham}, D.~W., \& {Mayor}, M. 2000, \apjl,
  529, L45

\bibitem[{{Charbonneau} {et~al.}(2002){Charbonneau}, {Brown}, {Noyes}, \&
  {Gilliland}}]{charbonneau_et_al2002}
{Charbonneau}, D., {Brown}, T.~M., {Noyes}, R.~W., \& {Gilliland}, R.~L. 2002,
  \apj, 568, 377

\bibitem[{{Cho} {et~al.}(2003){Cho}, {Menou}, {Hansen}, \&
  {Seager}}]{cho_et_al2003}
{Cho}, J.~Y.-K., {Menou}, K., {Hansen}, B.~M.~S., \& {Seager}, S. 2003, \apjl,
  587, L117

\bibitem[{{Collier Cameron} {et~al.}(2006){Collier Cameron}, {Pollacco},
  {Street}, {Lister}, {West}, {Wilson}, {Pont}, {Christian}, {Clarkson},
  {Enoch}, {Evans}, {Fitzsimmons}, {Haswell}, {Hellier}, {Hodgkin}, {Horne},
  {Irwin}, {Kane}, {Keenan}, {Norton}, {Parley}, {Osborne}, {Ryans}, {Skillen},
  \& {Wheatley}}]{collier_et_al2006}
{Collier Cameron}, A., {Pollacco}, D., {Street}, R.~A., {Lister}, T.~A.,
  {West}, R.~G., {Wilson}, D.~M., {Pont}, F., {Christian}, D.~J., {Clarkson},
  W.~I., {Enoch}, B., {Evans}, A., {Fitzsimmons}, A., {Haswell}, C.~A.,
  {Hellier}, C., {Hodgkin}, S.~T., {Horne}, K., {Irwin}, J., {Kane}, S.~R.,
  {Keenan}, F.~P., {Norton}, A.~J., {Parley}, N.~R., {Osborne}, J., {Ryans},
  R., {Skillen}, I., \& {Wheatley}, P.~J. 2006, \mnras, 373, 799

\bibitem[{{Cooper} \& {Showman}(2005)}]{cooper+showman2005}
{Cooper}, C.~S. \& {Showman}, A.~P. 2005, \apjl, 629, L45

\bibitem[{{Diego} {et~al.}(1995){Diego}, {Fish}, {Barlow}, {Crawford},
  {Spyromilio}, {Dryburgh}, {Brooks}, {Howarth}, \& {Walker}}]{diego_et_al1995}
{Diego}, F., {Fish}, A.~C., {Barlow}, M.~J., {Crawford}, I.~A., {Spyromilio},
  J., {Dryburgh}, M., {Brooks}, D., {Howarth}, I.~D., \& {Walker}, D.~D. 1995,
  \mnras, 272, 323

\bibitem[{{Dravins}(1985)}]{dravins1985}
{Dravins}, D. 1985, in IAU Colloq. 88: Stellar Radial Velocities, ed. A.~G.~D.
  {Philip} \& D.~W. {Latham}, 311--+

\bibitem[{{Fabrycky} {et~al.}(2007){Fabrycky}, {Johnson}, \&
  {Goodman}}]{fabrycky_et_al2007}
{Fabrycky}, D.~C., {Johnson}, E.~T., \& {Goodman}, J. 2007, ArXiv Astrophysics
  e-prints

\bibitem[{{Fortney}(2005)}]{fortney2005}
{Fortney}, J.~J. 2005, \mnras, 364, 649

\bibitem[{{Gaudi}(2005)}]{gaudi2005}
{Gaudi}, B.~S. 2005, \apjl, 628, L73

\bibitem[{{Gaudi} \& {Winn}(2007)}]{gaudi+winn2007}
{Gaudi}, B.~S. \& {Winn}, J.~N. 2007, \apj, 655, 550

\bibitem[{{Ge} {et~al.}(2002){Ge}, {Angel}, {Jacobsen}, {Woolf}, {Fugate},
  {Black}, \& {Lloyd-Hart}}]{ge_et_al2002}
{Ge}, J., {Angel}, J.~R.~P., {Jacobsen}, B., {Woolf}, N., {Fugate}, R.~Q.,
  {Black}, J.~H., \& {Lloyd-Hart}, M. 2002, \pasp, 114, 879

\bibitem[{{Ge} {et~al.}(2006){Ge}, {van Eyken}, {Mahadevan}, {DeWitt}, {Kane},
  {Cohen}, {Vanden Heuvel}, {Fleming}, {Guo}, {Henry}, {Schneider}, {Ramsey},
  {Wittenmyer}, {Endl}, {Cochran}, {Ford}, {Mart{\'{\i}}n}, {Israelian},
  {Valenti}, \& {Montes}}]{ge_et_al2006}
{Ge}, J., {van Eyken}, J., {Mahadevan}, S., {DeWitt}, C., {Kane}, S.~R.,
  {Cohen}, R., {Vanden Heuvel}, A., {Fleming}, S.~W., {Guo}, P., {Henry},
  G.~W., {Schneider}, D.~P., {Ramsey}, L.~W., {Wittenmyer}, R.~A., {Endl}, M.,
  {Cochran}, W.~D., {Ford}, E.~B., {Mart{\'{\i}}n}, E.~L., {Israelian}, G.,
  {Valenti}, J., \& {Montes}, D. 2006, \apj, 648, 683

\bibitem[{{Gillon} {et~al.}(2007){Gillon}, {Demory}, {Barman}, {Bonfils},
  {Mazeh}, {Udry}, {Mayor}, \& {Queloz}}]{gillon_et_al2007}
{Gillon}, M., {Demory}, B.~., {Barman}, T., {Bonfils}, X., {Mazeh}, T., {Udry},
  S., {Mayor}, M., \& {Queloz}, D. 2007, ArXiv e-prints, 707

\bibitem[{{Gim{\'e}nez}(2006)}]{gimenez_et_al2006}
{Gim{\'e}nez}, A. 2006, \apj, 650, 408

\bibitem[{{Goldreich} \& {Peale}(1966)}]{goldreich+peale1966}
{Goldreich}, P. \& {Peale}, S. 1966, \aj, 71, 425

\bibitem[{{Grillmair} {et~al.}(2007){Grillmair}, {Charbonneau}, {Burrows},
  {Armus}, {Stauffer}, {Meadows}, {Van Cleve}, \&
  {Levine}}]{grillmair_et_al2007}
{Grillmair}, C.~J., {Charbonneau}, D., {Burrows}, A., {Armus}, L., {Stauffer},
  J., {Meadows}, V., {Van Cleve}, J., \& {Levine}, D. 2007, \apjl, 658, L115

\bibitem[{{Guillot}(2005)}]{guillot2005}
{Guillot}, T. 2005, Annual Review of Earth and Planetary Sciences, 33, 493

\bibitem[{{Guillot} {et~al.}(1996){Guillot}, {Burrows}, {Hubbard}, {Lunine}, \&
  {Saumon}}]{guillot_et_al1996}
{Guillot}, T., {Burrows}, A., {Hubbard}, W.~B., {Lunine}, J.~I., \& {Saumon},
  D. 1996, \apjl, 459, L35+

\bibitem[{{Guillot} \& {Showman}(2002)}]{guillot+showman2002}
{Guillot}, T. \& {Showman}, A.~P. 2002, \aap, 385, 156

\bibitem[{{Harrington} {et~al.}(2006){Harrington}, {Hansen}, {Luszcz},
  {Seager}, {Deming}, {Menou}, {Cho}, \& {Richardson}}]{harrington_et_al2006}
{Harrington}, J., {Hansen}, B.~M., {Luszcz}, S.~H., {Seager}, S., {Deming}, D.,
  {Menou}, K., {Cho}, J.~Y.-K., \& {Richardson}, L.~J. 2006, Science, 314, 623

\bibitem[{{Henry} {et~al.}(2000){Henry}, {Marcy}, {Butler}, \&
  {Vogt}}]{henry_et_al2000}
{Henry}, G.~W., {Marcy}, G.~W., {Butler}, R.~P., \& {Vogt}, S.~S. 2000, \apjl,
  529, L41

\bibitem[{{Howard} {et~al.}(1984){Howard}, {Gilman}, \&
  {Gilman}}]{howard_et_al1984}
{Howard}, R., {Gilman}, P.~I., \& {Gilman}, P.~A. 1984, \apj, 283, 373

\bibitem[{{Hui} \& {Seager}(2002)}]{hui+seager2002}
{Hui}, L. \& {Seager}, S. 2002, \apj, 572, 540

\bibitem[{{Johns-Krull} {et~al.}(2007){Johns-Krull}, {McCullough}, {Burke},
  {Valenti}, {Janes}, {Heasley}, {Bissinger}, {Fleenor}, {Foote},
  {Garcia-Melendo}, {Gary}, {Howell}, {Mallia}, {Masi}, {Prato}, \&
  {Vanmunster}}]{johnskrull_et_al2007}
{Johns-Krull}, C.~M., {McCullough}, P.~M., {Burke}, C.~J., {Valenti}, J.~A.,
  {Janes}, K.~A., {Heasley}, J.~N., {Bissinger}, R., {Fleenor}, M., {Foote},
  C.~N., {Garcia-Melendo}, E., {Gary}, B.~L., {Howell}, P.~J., {Mallia}, F.,
  {Masi}, G., {Prato}, L.~A., \& {Vanmunster}, T. 2007, in American
  Astronomical Society Meeting Abstracts, Vol. 210, American Astronomical
  Society Meeting Abstracts, 96.05--+

\bibitem[{{Knutson} {et~al.}(2007{\natexlab{a}}){Knutson}, {Charbonneau},
  {Allen}, {Fortney}, {Agol}, {Cowan}, {Showman}, {Cooper}, \&
  {Megeath}}]{knutson_et_al2007b}
{Knutson}, H.~A., {Charbonneau}, D., {Allen}, L.~E., {Fortney}, J.~J., {Agol},
  E., {Cowan}, N.~B., {Showman}, A.~P., {Cooper}, C.~S., \& {Megeath}, S.~T.
  2007{\natexlab{a}}, \nat, 447, 183

\bibitem[{{Knutson} {et~al.}(2007{\natexlab{b}}){Knutson}, {Charbonneau},
  {Noyes}, {Brown}, \& {Gilliland}}]{knutson_et_al2007a}
{Knutson}, H.~A., {Charbonneau}, D., {Noyes}, R.~W., {Brown}, T.~M., \&
  {Gilliland}, R.~L. 2007{\natexlab{b}}, \apj, 655, 564

\bibitem[{{Konacki} {et~al.}(2003){Konacki}, {Torres}, {Jha}, \&
  {Sasselov}}]{konacki_et_al2003}
{Konacki}, M., {Torres}, G., {Jha}, S., \& {Sasselov}, D.~D. 2003, \nat, 421,
  507

\bibitem[{{Konacki} {et~al.}(2005){Konacki}, {Torres}, {Sasselov}, \&
  {Jha}}]{konacki_et_al2005}
{Konacki}, M., {Torres}, G., {Sasselov}, D.~D., \& {Jha}, S. 2005, \apj, 624,
  372

\bibitem[{{Konacki} {et~al.}(2004){Konacki}, {Torres}, {Sasselov},
  {Pietrzy{\'n}ski}, {Udalski}, {Jha}, {Ruiz}, {Gieren}, \&
  {Minniti}}]{konacki_et_al2004}
{Konacki}, M., {Torres}, G., {Sasselov}, D.~D., {Pietrzy{\'n}ski}, G.,
  {Udalski}, A., {Jha}, S., {Ruiz}, M.~T., {Gieren}, W., \& {Minniti}, D. 2004,
  \apjl, 609, L37

\bibitem[{{Laughlin} {et~al.}(2005){Laughlin}, {Wolf}, {Vanmunster},
  {Bodenheimer}, {Fischer}, {Marcy}, {Butler}, \& {Vogt}}]{laughlin_et_al2005}
{Laughlin}, G., {Wolf}, A., {Vanmunster}, T., {Bodenheimer}, P., {Fischer}, D.,
  {Marcy}, G., {Butler}, P., \& {Vogt}, S. 2005, \apj, 621, 1072

\bibitem[{{Levrard} {et~al.}(2007){Levrard}, {Correia}, {Chabrier}, {Baraffe},
  {Selsis}, \& {Laskar}}]{levrard_et_al2007}
{Levrard}, B., {Correia}, A.~C.~M., {Chabrier}, G., {Baraffe}, I., {Selsis},
  F., \& {Laskar}, J. 2007, \aap, 462, L5

\bibitem[{{Loeb}(2005)}]{loeb2005}
{Loeb}, A. 2005, \apjl, 623, L45

\bibitem[{{Mayor} \& {Queloz}(1995)}]{mayor+queloz1995}
{Mayor}, M. \& {Queloz}, D. 1995, \nat, 378, 355

\bibitem[{{McCullough} {et~al.}(2006){McCullough}, {Stys}, {Valenti},
  {Johns-Krull}, {Janes}, {Heasley}, {Bye}, {Dodd}, {Fleming}, {Pinnick},
  {Bissinger}, {Gary}, {Howell}, \& {Vanmunster}}]{mccullough_et_al2006}
{McCullough}, P.~R., {Stys}, J.~E., {Valenti}, J.~A., {Johns-Krull}, C.~M.,
  {Janes}, K.~A., {Heasley}, J.~N., {Bye}, B.~A., {Dodd}, C., {Fleming}, S.~W.,
  {Pinnick}, A., {Bissinger}, R., {Gary}, B.~L., {Howell}, P.~J., \&
  {Vanmunster}, T. 2006, \apj, 648, 1228

\bibitem[{{McLaughlin}(1924)}]{mclaughlin1924}
{McLaughlin}, D.~B. 1924, \apj, 60, 22

\bibitem[{{Narita} {et~al.}(2007){Narita}, {Enya}, {Sato}, {Ohta}, {Winn},
  {Suto}, {Taruya}, {Turner}, {Aoki}, {Tamura}, {Yamada}, \&
  {Yoshi}}]{narita_et_al2007}
{Narita}, N., {Enya}, K., {Sato}, B., {Ohta}, Y., {Winn}, J.~N., {Suto}, Y.,
  {Taruya}, A., {Turner}, E.~L., {Aoki}, W., {Tamura}, M., {Yamada}, T., \&
  {Yoshi}, Y. 2007, ArXiv Astrophysics e-prints

\bibitem[{{O'Donovan} {et~al.}(2006){O'Donovan}, {Charbonneau}, {Mandushev},
  {Dunham}, {Latham}, {Torres}, {Sozzetti}, {Brown}, {Trauger}, {Belmonte},
  {Rabus}, {Almenara}, {Alonso}, {Deeg}, {Esquerdo}, {Falco}, {Hillenbrand},
  {Roussanova}, {Stefanik}, \& {Winn}}]{odonovan_et_al2006b}
{O'Donovan}, F.~T., {Charbonneau}, D., {Mandushev}, G., {Dunham}, E.~W.,
  {Latham}, D.~W., {Torres}, G., {Sozzetti}, A., {Brown}, T.~M., {Trauger},
  J.~T., {Belmonte}, J.~A., {Rabus}, M., {Almenara}, J.~M., {Alonso}, R.,
  {Deeg}, H.~J., {Esquerdo}, G.~A., {Falco}, E.~E., {Hillenbrand}, L.~A.,
  {Roussanova}, A., {Stefanik}, R.~P., \& {Winn}, J.~N. 2006, \apjl, 651, L61

\bibitem[{{Ohta} {et~al.}(2005){Ohta}, {Taruya}, \& {Suto}}]{ohta_et_al2005}
{Ohta}, Y., {Taruya}, A., \& {Suto}, Y. 2005, \apj, 622, 1118

\bibitem[{{Ohta} {et~al.}(2006){Ohta}, {Taruya}, \& {Suto}}]{ohta_et_al2006}
---. 2006, ArXiv Astrophysics e-prints

\bibitem[{{Pont} {et~al.}(2004){Pont}, {Bouchy}, {Queloz}, {Santos}, {Melo},
  {Mayor}, \& {Udry}}]{pont_et_al2004}
{Pont}, F., {Bouchy}, F., {Queloz}, D., {Santos}, N.~C., {Melo}, C., {Mayor},
  M., \& {Udry}, S. 2004, \aap, 426, L15

\bibitem[{{Poynting}(1903)}]{poynting1903}
{Poynting}, J.~H. 1903, \mnras, 64, A1+

\bibitem[{{Press} \& {Rybicki}(1993)}]{press+rybicki1993}
{Press}, W.~H. \& {Rybicki}, G.~B. 1993, \apj, 418, 585

\bibitem[{{Richardson} {et~al.}(2007){Richardson}, {Deming}, {Horning},
  {Seager}, \& {Harrington}}]{richardson_et_al2007}
{Richardson}, L.~J., {Deming}, D., {Horning}, K., {Seager}, S., \&
  {Harrington}, J. 2007, \nat, 445, 892

\bibitem[{{Richardson} {et~al.}(2003){Richardson}, {Deming}, \&
  {Seager}}]{richardson_et_al2003}
{Richardson}, L.~J., {Deming}, D., \& {Seager}, S. 2003, \apj, 597, 581

\bibitem[{{Robertson}(1937)}]{robertson1937}
{Robertson}, H.~P. 1937, \mnras, 97, 423

\bibitem[{{Rossiter}(1924)}]{rossiter1924}
{Rossiter}, R.~A. 1924, \apj, 60, 15

\bibitem[{{Seager} \& {Hui}(2002)}]{seager+hui2002}
{Seager}, S. \& {Hui}, L. 2002, \apj, 574, 1004

\bibitem[{{Seager} \& {Sasselov}(2000)}]{seager+sasselov2000}
{Seager}, S. \& {Sasselov}, D.~D. 2000, \apj, 537, 916

\bibitem[{{Showman} \& {Guillot}(2002)}]{showman+guillot2002}
{Showman}, A.~P. \& {Guillot}, T. 2002, \aap, 385, 166

\bibitem[{{Silva}(2003)}]{silva2003}
{Silva}, A.~V.~R. 2003, \apjl, 585, L147

\bibitem[{{Sudarsky} {et~al.}(2003){Sudarsky}, {Burrows}, \&
  {Hubeny}}]{sudarsky_et_al2003}
{Sudarsky}, D., {Burrows}, A., \& {Hubeny}, I. 2003, \apj, 588, 1121

\bibitem[{{Sudarsky} {et~al.}(2000){Sudarsky}, {Burrows}, \&
  {Pinto}}]{sudarsky_et_al2000}
{Sudarsky}, D., {Burrows}, A., \& {Pinto}, P. 2000, \apj, 538, 885

\bibitem[{{Swain} {et~al.}(2007){Swain}, {Bouwman}, {Akeson}, {Lawler}, \&
  {Beichman}}]{swain_et_al2007}
{Swain}, M.~R., {Bouwman}, J., {Akeson}, R., {Lawler}, S., \& {Beichman}, C.
  2007, ArXiv Astrophysics e-prints

\bibitem[{{Tokunaga} {et~al.}(2006){Tokunaga}, {Bond}, {Elias}, {Chun},
  {Richter}, {Liang}, {Lacy}, {Daggert}, {Tollestrup}, {Ressler}, {Warren},
  {Fisher}, \& {Carr}}]{tokunaga_et_al2006}
{Tokunaga}, A.~T., {Bond}, T., {Elias}, J., {Chun}, M., {Richter}, M., {Liang},
  M., {Lacy}, J., {Daggert}, L., {Tollestrup}, E., {Ressler}, M., {Warren}, D.,
  {Fisher}, S., \& {Carr}, J. 2006, in Ground-based and Airborne
  Instrumentation for Astronomy. Edited by McLean, Ian S.; Iye, Masanori.
  Proceedings of the SPIE, Volume 6269, pp. 62693Y (2006).

\bibitem[{{Trilling} {et~al.}(2002){Trilling}, {Lunine}, \&
  {Benz}}]{trilling_et_al2002}
{Trilling}, D.~E., {Lunine}, J.~I., \& {Benz}, W. 2002, \aap, 394, 241

\bibitem[{{Udalski} {et~al.}(2003){Udalski}, {Pietrzynski}, {Szymanski},
  {Kubiak}, {Zebrun}, {Soszynski}, {Szewczyk}, \&
  {Wyrzykowski}}]{udalski_et_al2003}
{Udalski}, A., {Pietrzynski}, G., {Szymanski}, M., {Kubiak}, M., {Zebrun}, K.,
  {Soszynski}, I., {Szewczyk}, O., \& {Wyrzykowski}, L. 2003, Acta Astronomica,
  53, 133

\bibitem[{{Udalski} {et~al.}(2002{\natexlab{a}}){Udalski}, {Szewczyk},
  {Zebrun}, {Pietrzynski}, {Szymanski}, {Kubiak}, {Soszynski}, \&
  {Wyrzykowski}}]{udalski_et_al2002a}
{Udalski}, A., {Szewczyk}, O., {Zebrun}, K., {Pietrzynski}, G., {Szymanski},
  M., {Kubiak}, M., {Soszynski}, I., \& {Wyrzykowski}, L. 2002{\natexlab{a}},
  Acta Astronomica, 52, 317

\bibitem[{{Udalski} {et~al.}(2002{\natexlab{b}}){Udalski}, {Szymanski},
  {Kubiak}, {Pietrzynski}, {Soszynski}, {Wozniak}, {Zebrun}, {Szewczyk}, \&
  {Wyrzykowski}}]{udalski_et_al2002b}
{Udalski}, A., {Szymanski}, M., {Kubiak}, M., {Pietrzynski}, G., {Soszynski},
  I., {Wozniak}, P., {Zebrun}, K., {Szewczyk}, O., \& {Wyrzykowski}, L.
  2002{\natexlab{b}}, Acta Astronomica, 52, 217

\bibitem[{{Udalski} {et~al.}(2004){Udalski}, {Szymanski}, {Kubiak},
  {Pietrzynski}, {Soszynski}, {Zebrun}, {Szewczyk}, \&
  {Wyrzykowski}}]{udalski_et_al2004}
{Udalski}, A., {Szymanski}, M.~K., {Kubiak}, M., {Pietrzynski}, G.,
  {Soszynski}, I., {Zebrun}, K., {Szewczyk}, O., \& {Wyrzykowski}, L. 2004,
  Acta Astronomica, 54, 313

\bibitem[{{Udalski} {et~al.}(2002{\natexlab{c}}){Udalski}, {Zebrun},
  {Szymanski}, {Kubiak}, {Soszynski}, {Szewczyk}, {Wyrzykowski}, \&
  {Pietrzynski}}]{udalski_et_al2002c}
{Udalski}, A., {Zebrun}, K., {Szymanski}, M., {Kubiak}, M., {Soszynski}, I.,
  {Szewczyk}, O., {Wyrzykowski}, L., \& {Pietrzynski}, G. 2002{\natexlab{c}},
  Acta Astronomica, 52, 115

\bibitem[{{Ulrich}(1991)}]{ulrich1991}
{Ulrich}, R.~K. 1991, Advances in Space Research, 11, 217

\bibitem[{{Vidal-Madjar} {et~al.}(2004){Vidal-Madjar}, {D{\'e}sert},
  {Lecavelier des Etangs}, {H{\'e}brard}, {Ballester}, {Ehrenreich}, {Ferlet},
  {McConnell}, {Mayor}, \& {Parkinson}}]{vidal-madjar_et_al2004}
{Vidal-Madjar}, A., {D{\'e}sert}, J.-M., {Lecavelier des Etangs}, A.,
  {H{\'e}brard}, G., {Ballester}, G.~E., {Ehrenreich}, D., {Ferlet}, R.,
  {McConnell}, J.~C., {Mayor}, M., \& {Parkinson}, C.~D. 2004, \apjl, 604, L69

\bibitem[{{Vidal-Madjar} {et~al.}(2003){Vidal-Madjar}, {Lecavelier des Etangs},
  {D{\'e}sert}, {Ballester}, {Ferlet}, {H{\'e}brard}, \&
  {Mayor}}]{vidal-madjar_et_al2003}
{Vidal-Madjar}, A., {Lecavelier des Etangs}, A., {D{\'e}sert}, J.-M.,
  {Ballester}, G.~E., {Ferlet}, R., {H{\'e}brard}, G., \& {Mayor}, M. 2003,
  \nat, 422, 143

\bibitem[{{Winn} \& {Holman}(2005)}]{winn+holman2005}
{Winn}, J.~N. \& {Holman}, M.~J. 2005, \apjl, 628, L159

\bibitem[{{Winn} {et~al.}(2006){Winn}, {Johnson}, {Marcy}, {Butler}, {Vogt},
  {Henry}, {Roussanova}, {Holman}, {Enya}, {Narita}, {Suto}, \&
  {Turner}}]{winn_et_al2006}
{Winn}, J.~N., {Johnson}, J.~A., {Marcy}, G.~W., {Butler}, R.~P., {Vogt},
  S.~S., {Henry}, G.~W., {Roussanova}, A., {Holman}, M.~J., {Enya}, K.,
  {Narita}, N., {Suto}, Y., \& {Turner}, E.~L. 2006, \apjl, 653, L69

\bibitem[{{Winn} {et~al.}(2005){Winn}, {Noyes}, {Holman}, {Charbonneau},
  {Ohta}, {Taruya}, {Suto}, {Narita}, {Turner}, {Johnson}, {Marcy}, {Butler},
  \& {Vogt}}]{winn_et_al2005}
{Winn}, J.~N., {Noyes}, R.~W., {Holman}, M.~J., {Charbonneau}, D., {Ohta}, Y.,
  {Taruya}, A., {Suto}, Y., {Narita}, N., {Turner}, E.~L., {Johnson}, J.~A.,
  {Marcy}, G.~W., {Butler}, R.~P., \& {Vogt}, S.~S. 2005, \apj, 631, 1215

\end{thebibliography}

\end{document}